\documentclass[letterpaper,twocolumn,10pt]{article}
\usepackage{usenix2019_v3}

\newif\ifcameraready
\camerareadyfalse

\usepackage{authblk}

\usepackage{amsmath,amsfonts}
\usepackage[T1]{fontenc}
\usepackage[utf8]{inputenc}

\usepackage{upquote}

\usepackage{microtype}
\UseMicrotypeSet[protrusion]{basicmath}

\usepackage{tikz}
\usetikzlibrary{patterns}
\usetikzlibrary{patterns.meta}

\tikzdeclarepattern{
  name=mylines,
  parameters={
      \pgfkeysvalueof{/pgf/pattern keys/size},
      \pgfkeysvalueof{/pgf/pattern keys/angle},
      \pgfkeysvalueof{/pgf/pattern keys/line width},
  },
  bounding box={
    (0,-0.5*\pgfkeysvalueof{/pgf/pattern keys/line width}) and
    (\pgfkeysvalueof{/pgf/pattern keys/size},
0.5*\pgfkeysvalueof{/pgf/pattern keys/line width})},
  tile size={(\pgfkeysvalueof{/pgf/pattern keys/size},
\pgfkeysvalueof{/pgf/pattern keys/size})},
  tile transformation={rotate=\pgfkeysvalueof{/pgf/pattern keys/angle}},
  defaults={
    size/.initial=5pt,
    angle/.initial=45,
    line width/.initial=.4pt,
  },
  code={
      \draw [line width=\pgfkeysvalueof{/pgf/pattern keys/line width}]
        (0,0) -- (\pgfkeysvalueof{/pgf/pattern keys/size},0);
  },
}
\newlength{\hatchspread}
\newlength{\hatchthickness}
\newlength{\hatchshift}
\newcommand{\hatchcolor}{}
\tikzset{hatchspread/.code={\setlength{\hatchspread}{#1}},
         hatchthickness/.code={\setlength{\hatchthickness}{#1}},
         hatchshift/.code={\setlength{\hatchshift}{#1}},
         hatchcolor/.code={\renewcommand{\hatchcolor}{#1}}}
\tikzset{hatchspread=3pt,
         hatchthickness=0.4pt,
         hatchshift=0pt, % must >= 0pt
         hatchcolor=black}
\pgfdeclarepatternformonly[\hatchspread,\hatchthickness,\hatchshift,\hatchcolor]% variables
   {custom north west lines}% name
   {\pgfqpoint{\dimexpr-2\hatchthickness}{\dimexpr-2\hatchthickness}}% lower left corner
   {\pgfqpoint{\dimexpr\hatchspread+2\hatchthickness}{\dimexpr\hatchspread+2\hatchthickness}}% upper right corner
   {\pgfqpoint{\dimexpr\hatchspread}{\dimexpr\hatchspread}}% tile size
   {% shape description
    \pgfsetlinewidth{\hatchthickness}
    \pgfpathmoveto{\pgfqpoint{0pt}{\dimexpr\hatchspread+\hatchshift}}
    \pgfpathlineto{\pgfqpoint{\dimexpr\hatchspread+0.15pt+\hatchshift}{-0.15pt}}
    \ifdim \hatchshift > 0pt
      \pgfpathmoveto{\pgfqpoint{0pt}{\hatchshift}}
      \pgfpathlineto{\pgfqpoint{\dimexpr0.15pt+\hatchshift}{-0.15pt}}
    \fi
    \pgfsetstrokecolor{\hatchcolor}
    \pgfusepath{stroke}
   }
\pgfdeclarepatternformonly[\hatchspread,\hatchthickness,\hatchshift,\hatchcolor]% variables
   {custom north east lines}% name
   {\pgfqpoint{\dimexpr-2\hatchthickness}{\dimexpr-2\hatchthickness}}% lower left corner
   {\pgfqpoint{\dimexpr\hatchspread+2\hatchthickness}{\dimexpr\hatchspread+2\hatchthickness}}% upper right corner
   {\pgfqpoint{\dimexpr\hatchspread}{\dimexpr\hatchspread}}% tile size
   {% shape description
    \pgfsetlinewidth{\hatchthickness}
    \pgfpathmoveto{\pgfqpoint{\dimexpr\hatchshift-0.15pt}{-0.15pt}}
    \pgfpathlineto{\pgfqpoint{\dimexpr\hatchspread+0.15pt}{\dimexpr\hatchspread-\hatchshift+0.15pt}}
    \ifdim \hatchshift > 0pt
      \pgfpathmoveto{\pgfqpoint{-0.15pt}{\dimexpr\hatchspread-\hatchshift-0.15pt}}
      \pgfpathlineto{\pgfqpoint{\dimexpr\hatchshift+0.15pt}{\dimexpr\hatchspread+0.15pt}}
    \fi
    \pgfsetstrokecolor{\hatchcolor}
    \pgfusepath{stroke}
   }

\usetikzlibrary{positioning}
\usetikzlibrary{shapes.geometric}
\usetikzlibrary{shapes.multipart}
\usetikzlibrary{shapes.misc}
\usetikzlibrary{arrows}
\usetikzlibrary{arrows.meta}
\usepackage{pgfplots}
\usepackage{pgfplotstable}
\usepgfplotslibrary{fillbetween}
\usetikzlibrary{pgfplots.groupplots}
\pgfplotsset{compat=1.15}

\usepackage{hyperref}
\usepackage{graphicx,grffile,array,xcolor}
\usetikzlibrary{matrix}

\usepackage{times} 
\usepackage{multirow}
\usepackage{xspace}
\usepackage{tabularx}
\usepackage{booktabs}
\usepackage{paralist}
\usepackage[american]{babel}
\usepackage[shortlabels]{enumitem}
\usepackage{courier,wrapfig}
\usepackage{xspace}
\usepackage{balance}
\usepackage{enumitem}
\usepackage{epstopdf}
\usepackage{multirow}
\usepackage{booktabs}
\usepackage{url}
\usepackage{pifont}
\usepackage{subfig}
\usepackage{comment}
\usepackage{mathtools}
\usepackage[normalem]{ulem}
\usepackage{xkeyval}

\usepackage{algorithm}
\usepackage{algpseudocode}
\usepackage{etoolbox}
\usepackage{ifthen}
\usepackage{sepfootnotes}
\usepackage[compact]{titlesec}

\usepackage[colorinlistoftodos]{todonotes} % enable

\usepackage{pifont}
\usepackage{caption}
\DeclareCaptionFormat{myformat}{#1#2#3\hrulefill}
\usepackage[export]{adjustbox}

\PassOptionsToPackage{usenames,dvipsnames}{color} % color is loaded by
\hypersetup{unicode=true,
            colorlinks=true,
            linkcolor=green,
            citecolor=red,
            anchorcolor=pink,
            urlcolor=blue,
            breaklinks=true}
\makeatletter
\def\maxwidth{\ifdim\Gin@nat@width>\linewidth\linewidth\else\Gin@nat@width\fi}
\def\maxheight{\ifdim\Gin@nat@height>\textheight\textheight\else\Gin@nat@height\fi}
\makeatother
\setkeys{Gin}{width=\maxwidth,height=\maxheight,keepaspectratio}
\makeatletter
\g@addto@macro{\UrlBreaks}{\UrlOrds}
\makeatother

\apptocmd\normalsize{%
\abovedisplayskip=5pt
\abovedisplayshortskip=5pt
\belowdisplayskip=5pt
\belowdisplayshortskip=5pt
}{}{}

\newcommand{\sysname}{NetBlaze\xspace}

\newcommand{\pninetynine}{p99\xspace}
\newcommand{\pninetyfive}{p95\xspace}

\newcommand{\pfifty}{p50\xspace}

\newcommand\paraspace{\vspace*{0.5ex}}
\newcommand\parab[1]{\paraspace\noindent\textbf{#1}}
\newcommand\parae[1]{\paraspace\textbf{\textit{#1}}}

\newcommand{\tuple}[2]{\langle #1,#2 \rangle}

\newcommand{\ie}{\emph{i.e.,}\xspace}
\newcommand{\eg}{\emph{e.g.,}\xspace}

\newcommand{\secref}[1]{\S\ref{#1}}
\newcommand{\figref}[1]{Fig.~\ref{#1}}
\newcommand{\tabref}[1]{Table~\ref{#1}}

\ifcameraready

\fi

\ifcameraready
\fi

\begin{document}

\date{}

% \title{\sysname: Microsecond-Scale Latency SLOs For Network Function Chains}

\title{Scheduling Network Function Chains Under Sub-Millisecond Latency SLOs}

% \author{Paper \#57}

% usenix format
% \author{
% {\rm Jianfeng Wang}\\
% University of Southern\\ California\\
% % jianfenw@usc.edu\\
% \and
% \rm{Siddhant Gupta}\\
% University of Southern\\ California\\
% % sgupta86@usc.edu\\
% \and
% \rm{Marcos A. M. Vieira}\\
% Universidade Federal de \quad\\Minas Gerais\\
% % mmvieira@dcc.ufmg.br\\
% \and
% \rm{Ramesh Govindan}\\
% University of Southern\\ California\\
% % ramesh@usc.edu\\
% \and
% \quad \rm{Barath Raghavan}\\
% \quad University of Southern\\ California\\
% % barathra@usc.edu\\
% }

\author[1]{Jianfeng Wang}
\author[1]{Siddhant Gupta}
\author[2]{Marcos A. M. Vieira}
\author[1]{Barath Raghavan}
\author[1]{Ramesh Govindan}

% \affil[ ]{\textit{contact: {jianfenw@usc.edu}}
\affil[1]{\textit{University of Southern California}}
\affil[2]{\textit{Universidade Federal de Minas Gerais}}

\maketitle

\begin{abstract}

% Network Function Virtualization (NFV) replaces hardware middleboxes with software-based Network Functions (NFs).
% % Past efforts focused on performance improvements under latency SLOs.
% To date, no NFV systems efficiently meet microsecond-scale latency SLOs. Prior work has led to the belief that for a service chain, every CPU core must prepare for worst-case delay.
% Thus, a large number of cores need to be provisioned to ensure low latency, especially as the target latency approaches the microsecond scale.

% Surprisingly, our analysis on real-world traces shows that today's auto-scaling / load-balancing mechanisms cannot meet microsecond-scale tail latency SLOs even with unlimited CPU cores, due to two types of transient traffic bursts.
% We present \sysname, an efficient and timely NFV scheduler that detects and processes bursts.
% \sysname schedules NF tasks to control queueing effects at hardware and software queues with packet-level traffic statistics.
% To achieve high efficiency, \sysname supports fine-grained core reallocation, and introduces sidecar cores for handling transient bursts rather than over-provisioning.
% Our evaluation shows that: compared to state-of-the-art NFV auto-scaling and load-balancing approaches, \sysname is able to meet 100s-$\mu$s latency SLOs  (up to 30$\times$ improvements) while using X\% less time-averaged CPU cores.

Network Function Virtualization (NFV) seeks to replace hardware middleboxes with software-based Network Functions (NFs). NFV systems are seeing greater deployment in the cloud and at the edge. However, especially at the edge, there is a mismatch between the traditional focus on NFV throughput and the need to meet very low latency SLOs, as edge services inherently require low latency. Moreover, cloud-based NFV systems need to achieve such low latency while minimizing CPU core usage.

We find that real-world traffic exhibits burstiness that causes latency spikes of up to 10s of milliseconds in existing NFV systems. To address this, we built \sysname, which achieves sub-millisecond p99 latency SLOs, even for adversarial traffic, using a novel multi-scale core-scaling strategy. \sysname makes traffic-to-core allocation decisions at rack, server, and core-spatial scales, and at increasingly finer timescales, to accommodate multi-timescale bursts. In comparison with state-of-the-art approaches, \sysname is the only one capable of achieving sub-millisecond p99 latency SLOs while using a comparable number of cores.

% They employ optimizations for high per-core performance, and scale up and down the number of CPU cores to adapt to the changing traffic volume.

% fast user-level networking, polling and run-to-completion execution to achieve high per-core performance, and scale up and down the number of cores used by each service chain with flow-level load balancing schemes.

% NFV systems need to deploy service chains with microsecond-scale tail latencies, as datacenter applications have advanced to the microsecond-scale era. A significant number of cores is required to host NF chains under low-latency SLOs.

% Many NFV systems use hardware and software queues to process dynamic traffic, and devote many CPU cores to each service chain to control the queueing delay at the tail to meet its latency SLO.
\end{abstract}
%%% Local Variables:
%%% mode: latex
%%% TeX-master: t
%%% End:

\section{Introduction}
\label{sec:intro}

Network Function Virtualization (NFV) enables processing packets in software, often using a chain of Network Functions (NFs). NFs can filter traffic, encrypt it, modify headers or payload in transit, and so on. A long line of work has shown that NFV improves network manageability over hardware middleboxes~~\cite{sigcomm14-making,sosp15-e2,osdi16-netbricks,sigcomm17-nfp,sigcomm17-nfvnice,nsdi18-resq,socc22-quadrant}. More recent work has demonstrated that NFV is performant~---~on commodity hardware, NFV systems can process complex NF chains at line rates~\cite{nsdi18-metron,conext19-rsspp,conext20-lemur,socc20-snf,socc22-quadrant}.

Spurred by these developments, cloud providers have started to offer NFV capabilities to customers. This is most visible at the edge of today's distributed clouds~\cite{gcp-distributed-cloud,aws-distributed-cloud,azure-distributed-cluster}, where cloud racks expose to customers the ability to process packets. In this setting, the latency of packet processing must be as small as possible. \textit{Execution of NF chains is pure overhead}, and at the edge where network latencies are low (in nearly 50 countries, RTTs to cloud are less than 20~ms~\cite{hotnets2020:pruning-edge-research}), even a few milliseconds latency can represent significant overhead, especially since increased latency can adversely impact revenue.

As processor speeds increase, many NF chains can process a single packet within tens of microseconds. Yet, today's NFV systems exhibit 99th percentile (\pninetynine) latencies on the order of several milliseconds (\secref{sec:evaluation}). As we describe below, much of the increased latency comes from queueing. In this paper, we ask: is it possible to design an NFV system that offers a microsecond-scale \pninetynine latency service-level objective (SLO)? 

A primary challenge we face is the \textit{flow-affinity} requirement for NFV (\secref{s:background}). NFV systems assign flows to cores, so that all packets in a flow are processed by the same core. This avoids expensive cross-core synchronization for \textit{stateful} NFs which update shared state on every packet. To ensure low latency while preserving flow-affinity, NFV systems \textit{auto-scale} core allocations~\cite{nsdi18-metron,socc20-snf,socc22-quadrant}~---~as traffic increases or decreases, they scale up or down (respectively) the number of cores allocated to traffic. This ensures that traffic can be processed in a timely manner, while at the same time minimizing core usage. The latter is especially important in the cloud context, where cores not used for packet processing can be used to process revenue-generating application workloads.

Auto-scaling algorithms in today's NFV systems fail to react to two kinds of bursts seen in real traffic that can significantly increase queueing latency (\secref{s:challenge}). The first results from instantaneous packet rates in a flow that exceed the processing capacity of a single core. The second is a burst of a large number of flow arrivals that also exceeds core capacity, since each new flow requires additional setup overhead for stateful NFs. These bursts can be short-lived (10~ms or shorter) and can impose queueing latencies of several milliseconds (\secref{s:challenge}).

Existing NFV auto-scaling approaches fail to achieve low latency in real traffic for three reasons (\secref{s:approaches}): they make core allocation decisions based on observed packet rates alone, so cannot account for flow arrival bursts; they use time-averaged estimates that cannot capture short-timescale bursts; they make core allocation decisions at longer timescales (of 100~ms to 1~s) than the duration of bursts in real traffic.

\parab{Contributions.} This paper describes the design and implementation of \sysname, a rack-scale NFV system that achieves microsecond-scale latency SLOs while ensuring high CPU efficiency (\secref{sec:design}). Its contributions include:
\begin{itemize}[nosep,leftmargin=*]
\item A novel hierarchical multi-scale flow-to-core mapping strategy that distributes core allocations across different spatial scales (ingress, server, core) and different temporal scales (seconds at ingress to microseconds at core (\secref{subsec:system-overview}). This ensures that \sysname can adapt to traffic changes at different timescales, while ensuring scalable core allocation decisions by making coarse-grained decisions at the ingress (which sees the most traffic) and fine-grained decisions at the core (which sees relatively lower traffic).
\item A more accurate core capacity estimation technique for NFs that takes instantaneous flow counts as well as packet rates into account.
\item A lightweight mechanism to recruit \textit{auxiliary} cores to handle bursts that can potentially violate latency SLOs.
\item A fast algorithm to remap RSS buckets at a server to core to ensure that no core experiences sustained overload.
\end{itemize}

Experiments using traffic traces from a backbone as well as from within an ISP (\secref{sec:evaluation}) show that \sysname can support \pninetynine latency SLOs on the order of 100-500~$\mu$s while the state-of-the-art NFV systems~\cite{nsdi18-metron,socc22-quadrant,infocom22-dyssect} exhibit 10$\times$ higher \pninetynine latency SLOs. Moreover, \sysname's CPU-efficiency is comparable to, or better than, all of these systems. An ablation study shows that each of its design decisions contributes significantly to \sysname's performance.

% %% \input{motivation} 

\section{Background, Motivation and Approach}
\label{sec:motivation}

To motivate the problem that \sysname addresses and the approach it takes, we begin with some background.

\subsection{Background}
\label{s:background}

%% Deployment
%% NFs and NF chains, latency SLOs
%% Core usage assumptions
%% How current systems auto-scale: maybe goes after section below

\parab{Deployment Scenario.} Many cloud providers use \textit{distributed cloud} designs with rack deployments at the edge of the network~\cite{gcp-distributed-cloud,aws-distributed-cloud,azure-distributed-cluster}. These rack deployments are either within the premises of an enterprise, or at the ingress of a carrier (\eg a telco). Each rack contains a ToR switch and several multi-core servers. Each server core can run edge applications, such as those envisioned for multi-access edge computing (MEC). Cores can
also run software NFs: indeed, cloud provider software in these deployments explicitly support software NFs. This paper targets NF processing in these deployments.

% \ramesh{Need to add references to GDC/E and Azure and Amazon offerings.}

\parab{NFs and NF chains.} NFs may process packets in a variety of ways: translate network addresses (\eg NAT), filter traffic (\eg using ACLs), encrypt traffic, and so forth. Usually, a flow is processed by an NF chain. A single edge rack may execute multiple NF chains to serve different classes of traffic. For each NF chain, an operator defines which traffic class (or traffic aggregate) the chain should process. 

\parae{Latency SLO.}
In the edge rack setting, it is desirable to bound the latency incurred in processing a packet using an NF chain. This enables cloud providers to meet end-to-end latency targets for applications. This bound, called a latency SLO, is expressed in terms of a high percentile (\eg the \pninetyfive or \pninetynine percentile, denoted \pninetyfive or \pninetynine) of the latency experienced by a packet traversing the NF chain. This latency consists of two primary parts: the time taken by the NF chain to process a packet, and the queueing delay experienced by packets.

\parae{Flow Affinity.} Many NFs are \textit{stateful}: they create and update internal state when processing packets. Most stateful NFs maintain per-flow state~\cite{nsdi17-stateless}: an NF creates state when it encounters a new flow, and every packet of the flow can potentially update the per-flow state. If packets of a flow are processed by different cores, NFs can incur significant synchronization and cache coherence overhead to ensure consistent updates to the per-flow state across cores. For this reason, NF systems usually preserve \textit{flow-affinity}: to the extent possible, packets of a flow are processed by a single core.

% what do we want to discuss in the motivation section?

% Our main focus is to find the set of mechanisms that is key for achieving high efficiency while meeting tail latency SLOs.
\parab{Auto-scaling.}
NFV systems like Metron~\cite{nsdi18-metron}, Quadrant~\cite{socc22-quadrant}, and Dyssect~\cite{infocom22-dyssect} assign NFs or NF chains to individual cores. In the settings that these papers consider, as in ours, the volume of traffic processed by an NF chain generally exceeds what a single core can handle, so each NF or NF chain is allocated more than one core.

These systems \textit{auto-scale} CPU core allocations: they dynamically change the number of cores assigned to an NF or NF chain. Specifically, auto-scaling (a) dynamically tracks increases and decreases in traffic and (b) correspondingly increases or decreases the number of cores assigned to each NF chain, with the aim of using the fewest cores necessary to process the traffic for that NF chain. This permits CPU-efficiency, since cores not used for packet processing can be used for application or background tasks.\footnote{Recent work~\cite{sosp17-zygos,nsdi19-shenango,osdi20-caladan,osdi20-racksched} makes a similar observation in the context of core allocation for RPC processing in clouds.}

These systems auto-scale core usage with different goals. Metron~\cite{nsdi18-metron} auto-scales to ensure enough cores to handle the current traffic, while minimizing packet losses due to any overloaded cores. Quadrant~\cite{socc22-quadrant} auto-scales to ensure a target \pninetynine latency SLO, while Dyssect~\cite{infocom22-dyssect} auto-scales to ensure a target \pfifty median latency. Our work focuses on auto-scaling for ensuring \pninetynine latency SLOs on the order of 200-300 $\mu$s.\footnote{Quadrant achieves similar SLOs, but on synthetic, non-bursty traffic. In \secref{sec:evaluation}, we show that, on real traces, its \pninetynine latencies can be as high as several milliseconds.}

% In this paper, we focus on \textit{auto-scaling}.
% An NFV system uses the auto-scaling mechanism to adapt to dynamic traffic.
% Auto-scaling are responsible for:
% 1) determining the set of active CPU cores in the cluster and distributing traffic among them;
% 2) detecting overloaded CPU cores and migrating excessive traffic from these cores to prevent latency SLO violations.

% Before we discuss the detailed system design, we introduce several key observations drawn from NFV systems when hosting stateful \nfchains to several realistic traffic.
% These observations are crucial to understand the often-ignored challenges in achieving both low latency and high efficiency for NFV systems.

\subsection{Challenges}
\label{s:challenge}

Two factors make auto-scaling challenging. The first is the latency SLO. If an NFV system allocates too few cores, packets can experience queueing, and significant SLO violations can occur. If it allocates too many, CPU efficiency can suffer. The second is flow-affinity: an auto-scaling solution must not violate flow-affinity; doing so would complicate NF design and incur significant performance penalties.\footnote{A large body of recent work has explored core allocation for low-overhead microsecond-scale RPCs~\cite{sosp17-zygos,nsdi19-shenango,osdi20-caladan,osdi20-racksched}.} The flow affinity constraint makes NF chain core allocation fundamentally different, so general task scheduling schemes like Join-the-Shortest-Queue (JSQ) scheduling do not apply in our context.

Beyond this, bursty traffic presents a challenge for efficient auto-scaling. When a traffic burst arrives, if an NF chain does not have an adequate number of cores, packets in the burst can encounter a queueing delay. If the burst is small, or if the latency SLO is loose, SLO violations may not occur. However, we find these extreme traffic cases do exist in real-world traffic and they cause significant SLO violations.

\parab{Superbursts.} In real packet traces, two types of large bursts (\textit{superbursts}) can occur that can result in SLO violations. To our knowledge, we are the first to identify superbursts as a challenge for core allocation in NF systems.

\parae{Whales.} A whale is (a part of) a \textit{single} flow whose packet processing requirements exceed the capacity of a single core. For example, suppose an NF-chain requires 10 $\mu$s to process a packet; if more than 10$^{3}$ packets arrive on a single flow within 10~ms, this would exceed the core's processing capacity during that time window. Whales arise because of the flow-affinity constraint in NF systems.

Packet traces captured in the wild contain whales. To show this, we use two packet traces: a city's backbone trace~\cite{caida-trace}, and one collected in a large autonomous system (AS)~\cite{as-trace}. For these two, we computed the \pninetynine latency resulting from \textit{processing each flow on a separate core} using 3 different types of NF chains: a \textit{light} chain that requires 5~$\mu$s of processing and \textit{medium} and \textit{heavy} chains that require 10 and 20 $\mu$s respectively. To do this, we use a packet-level discrete simulation framework. It simulates a rack-scale NFV system, including one ToR switch and many multi-core servers. It takes a realistic traffic trace as input, replays the trace and directs traffic to the simulated cluster running NF chains.

\begin{table}[t!]
\small
\centering
    \begin{tabular}{|m{2.55cm}|>{\centering\arraybackslash}m{1.6cm}|>{\centering\arraybackslash}m{1.6cm}|}
    \hline
    \multirow{2}{*}{\textbf{NF Chain}} & \multicolumn{2}{c|}{\textbf{Traffic Input}} \\
    \cline{2-3}
    & \textbf{Backbone} & \textbf{AS} \\
    \hline\hline
    \textbf{Light (5~$\mu$s)} 	& 831.3        & 53.8    \\ \hline
    \textbf{Medium (10~$\mu$s)}	& 2,690       & 578    \\ \hline
    \textbf{Heavy (20~$\mu$s)}  & 6,648      & 3,584  \\ \hline
\end{tabular}
\caption{Whales can inflate \pninetynine latency. Each cell represents the \pninetynine latency for the corresponding NF chain and trace in $\mu$s.}
\label{tab:whales}
\end{table}

In this setup, since every flow is processed by a single core, one expects the \pninetynine latency to be close to the NF chain's processing time. However, we find the overall \pninetynine latency to be more than two orders of magnitude higher in some cases (\tabref{tab:whales}): 6.6~ms for the backbone trace and 3.6~ms for the AS trace for the heavy chain. This indicates that in both those traces, one or more flows exhibit packet arrivals that, in the short-term, exceed core capacity and cause significant queueing that affects the overall latency.
% Under the city backbone trace, latency results are higher because 

\begin{table}[t!]
\small
\centering
    \begin{tabular}{|m{2.55cm}|>{\centering\arraybackslash}m{1.6cm}|>{\centering\arraybackslash}m{1.6cm}|}
    \hline
    \multirow{2}{*}{\textbf{NF Chain}} & \multicolumn{2}{c|}{\textbf{Traffic Input}} \\
    \cline{2-3}
    & \textbf{Backbone} & \textbf{AS} \\
    \hline\hline
    \textbf{Light (5~$\mu$s)} 	 & 4,029     & 3,062    \\ \hline
    \textbf{Medium (10~$\mu$s)}  & 3,830     & 14,241    \\ \hline
    \textbf{Heavy (20~$\mu$s)}   & 4,671     & 14,558  \\ \hline
\end{tabular}
\caption{Minnows can inflate \pninetynine latency. Each cell represents the \pninetynine latency for the corresponding NF chain and trace in $\mu$s.}
\label{tab:minnows}
\end{table}

\parae{Minnows.} Minnows represent a large number of active flows at a single core whose \textit{collective} processing requirements exceed the capacity of a single core. During auto-scaling, minnows can cause tail latency SLO violations even in the absence of whales. This is because: (a) if the corresponding NF chain is  stateful, the number of flows (possibly bursty ones) with packet arrivals within a short time period can be significantly larger than the median; (b) the state creation overhead for each flow can overwhelm core capacity. 
% Moreover, we show that processing more active flows at a core can result in high latency even under the same traffic input (\secref{}).

Real packet traces also contain minnows. Our methodology to demonstrate this uses the same two traces as described above, but: (a) uses flow-hashing (similar to these used in RSS-based systems) to distribute flows to cores, (b) assigns as many cores as necessary, so that the p50 latency for the entire trace is less than 10x the NF processing time and (c) shapes each flow, so that any burstiness is only due to flow arrivals, not packet arrivals.

Ideally, because traffic is shaped, one would expect the \pninetynine latency to be close to the \pfifty latency. However, as \tabref{tab:minnows} shows, the \pninetynine latency is much higher (nearly 14.6~ms for the heavy chain on the AS trace).

\subsection{Approach}
\label{s:approaches}

\parab{Shortcomings of existing approaches.} None of the existing systems~\cite{socc22-quadrant,nsdi18-metron,infocom22-dyssect} can maintain low \pninetynine latency SLOs in the presence of superbursts. We demonstrate this, in \secref{sec:evaluation}, using experiments on a cluster testbed using real traces~\footnote{Among them, Quadrant~\cite{socc22-quadrant} demonstrates p99 latency SLOs in the 50-300 $\mu$s range. However, their experiments replayed synthetic non-bursty traffic; we have verified this with the authors.}, but intuitively this is because:
\begin{itemize}[nosep,leftmargin=*]
\item They use only packet counts to make core re-allocation decisions. This does not consider minnow superbursts.
\item Some of them make core reallocation decisions at the traffic ingress using smoothed traffic statistics collected from individual servers. Smoothed traffic statistics are not sufficient to capture traffic demand shifts in short timescales. This also introduces new sources of delay in the measurement-reaction loop, which prevents the ingress from reacting quickly to transient bursts at individual CPU cores.
\item All of them make core reallocation decisions at a relatively coarse timescale (of 100~ms to 1~s). This is clearly not enough if the \pninetynine latency SLO is, say 500~$\mu$s. To meet the latter SLO, the system would need to detect and react\footnote{We cannot predict when minnows or whales occur. Proactive approaches such as splitting flows across cores (for whales) or conservatively allocating cores can result in inefficient CPU usage and/or increased processing overhead. Thus, we focus on \textit{reactive} strategies: detecting whales and minnows as quickly as possible, and quickly make core re-allocation decisions.} to potential latency violations at a finer timescale than 500~$\mu$s, or conservatively over-allocate cores across longer time intervals, sacrificing CPU efficiency.
\end{itemize}

\parab{\sysname's Approach.} \sysname permits \pninetynine latency SLOs on the order of 100~$\mu$s for traffic containing superbursts while still ensuring high CPU efficiency. It achieves this using two key ideas: hierarchical multi-scale allocation at different spatial and temporal scales, and using flow counts in addition to packet counts or rates to estimate core capacity. We describe these in the next section.

\section{\sysname Design}
\label{sec:design}

% 0) system overview and components:
% - NFVCtrl, NFVCore & NFVRCore, NFVMonitor (for generating NF profiles)
% Need to mention: why these components are new and critical for our design

This section begins with a high-level overview of \sysname before describing its algorithms in detail.

\subsection{Overview}
\label{subsec:system-overview}

\parab{Goal.} \sysname is a rack-scale NFV orchestration system that simultaneously seeks to achieve high CPU-efficiency (defined in terms of core-hours used) while ensuring microsecond-scale \pninetynine latency SLOs.

\parab{Assumptions.}
(1) Users of \sysname (\eg network operators) specify a set of NF chains to run on the rack. Each chain handles a traffic aggregate (\ie a group of flows) defined by the operator, who also specifies a \pninetynine latency SLO for the chain. \sysname aims to satisfy this SLO for flows processed by the NF chain. (2) Traffic aggregates are large enough that \sysname can dedicate a server to handle a chain; \sysname targets the enterprise edge or ISP ingress, where traffic volumes can be significant. (3) In most cases, an NF chain runs-to-completion~\cite{socc22-quadrant} on a core; run-to-completion has been shown to be a low overhead execution strategy for NF chains. When a core doesn't run NF chains, it can execute edge applications.

\begin{figure}[t]
\resizebox{\columnwidth}{!}{
\includegraphics[width=\columnwidth]{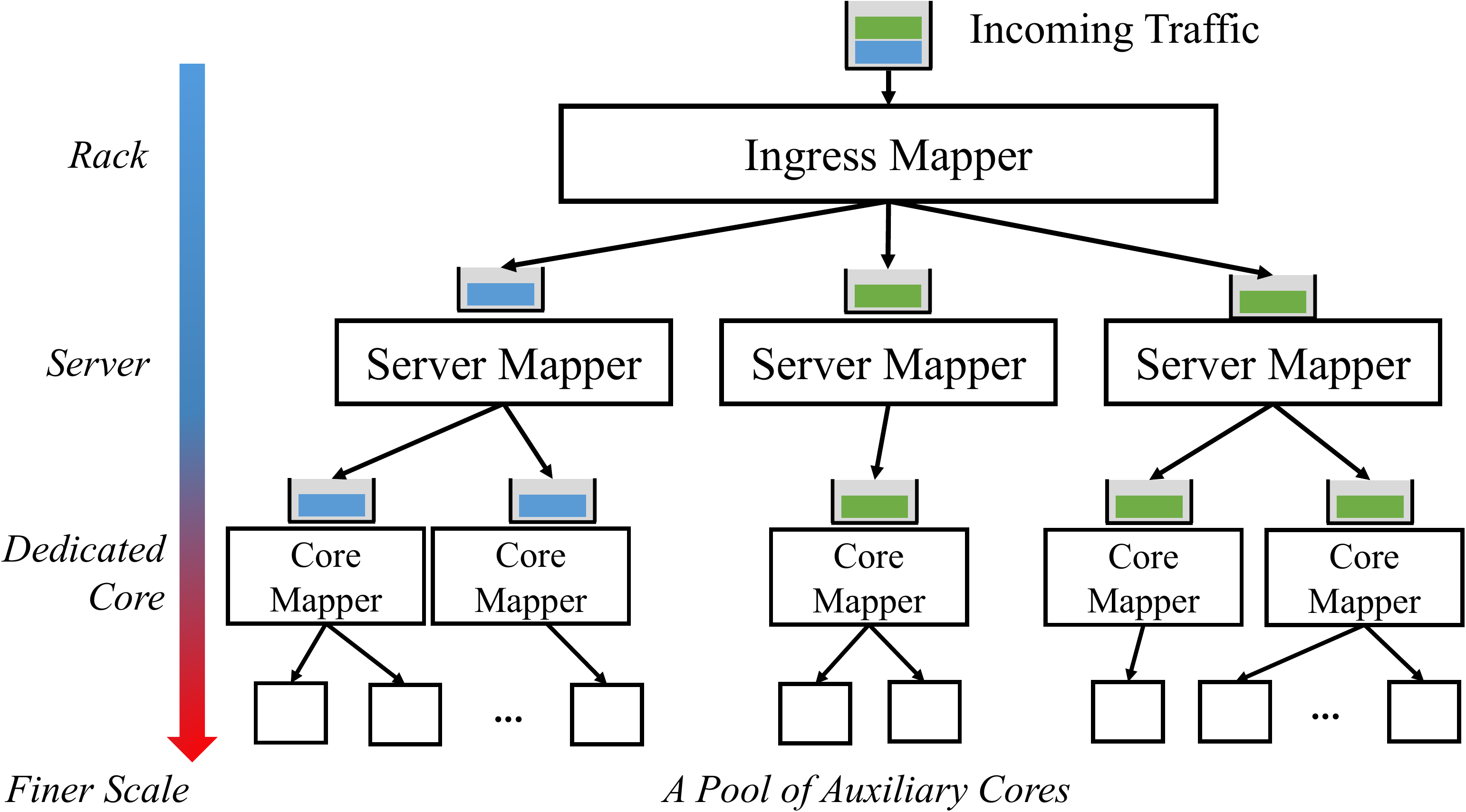}
}
\caption{\sysname's hierarchical multi-scale allocation maps traffic to cores at three different spatial scales and temporal scales.}
\label{fig:ironside-mappers}
\end{figure}

\parab{The Primary Challenge.} Given the flow-affinity constraint for NF processing, the primary challenge in the design of \sysname is to determine a \textit{policy} (or set of policies), together with associated mechanisms, for mapping flows to cores such that \sysname can achieve the goal described above. As discussed in \secref{s:approaches}, purely ingress-based allocation, using packet statistics, at the timescale of 100s of milliseconds, cannot work for \sysname. This begs the question: where should allocation decisions be made, and at what timescales?

\parab{Hierarchical Multi-scale Allocation.} The key idea in \sysname is to map flows to cores at three different spatial scales (ingress, server, and core), and at three different temporal scales (seconds, 100s of milliseconds, and 100s of $\mu$s, respectively), each with different objectives (\figref{fig:ironside-mappers}):
\begin{itemize}[nosep,leftmargin=*]
\item At an \textit{ingress mapper}, \sysname aims to minimize the number of servers dedicated to an NF chain.
\item \sysname's \textit{server mapper}, which runs on each \sysname server running NF chains. It attempts to use the fewest cores with high utilization, allowing \sysname to handle or adapt to sustained 100s of millisecond timescale bursts. It uses \textit{dedicated} cores for NF processing.
\item At each core, \sysname's \textit{core mapper} reactively mitigates $\mu$s-scale bursts by recruiting \textit{auxiliary} cores from a pool. An auxiliary core is generally used for other applications, but can be temporarily recruited for NF chain processing. %The key idea is that when allocating resources, each \sysname worker maintains a pool of auxiliary cores that can be temporarily used by dedicated cores to process bursts.
\end{itemize}

% \ramesh{Add a figure showing this?}

\parae{Handling minnows and whales.} The core mapper, when it detects potential SLO violations due to minnows, rapidly recruits one or more auxiliary cores (with the same server) and \textit{migrates} flows to them. When the core mapper detects a whale, it \textit{splits} NFs in the NF chain, executing one or more NFs on an auxiliary core. Once auxiliary cores finish processing the assigned flows, they resume application processing.

\parae{Dealing with flow count dependence.} To map flows to cores, NFV systems need to estimate or predict, for an NF chain, how many packets or flows a chain can handle for a given latency SLO. Some systems determine this dynamically and reactively~\cite{nsdi18-metron}, others use \textit{core capacity predictors} derived using offline profiling~\cite{socc22-quadrant}. Given \sysname's stringent time requirements, it uses the latter approach. However, unlike prior systems whose core capacity predictors are based on packet counts, \sysname's predicts core capacity in terms of the flow count \textit{and} the packet count, since high flow counts can violate latency SLOs (\secref{s:challenge}).

In the following subsections, we describe hierarchical multi-scale allocation in more detail. To simplify the description, we focus on a single NF chain hosted on the rack. Our \sysname implementation supports, using straightforward extensions, multiple NF chains running on the rack.

\subsection{Absorbing bursts: The Core Mapper}
\label{s:absorb-bursts:-core}

%% Goal: prevent SLO violations from bursty traffic
%% Detecting potential violations
%% Capacity prediction
%% Migration policy

\begin{figure}[t]
% \centering
% \resizebox{1.7\columnwidth}{!}{
\centering
\includegraphics[width=\columnwidth]{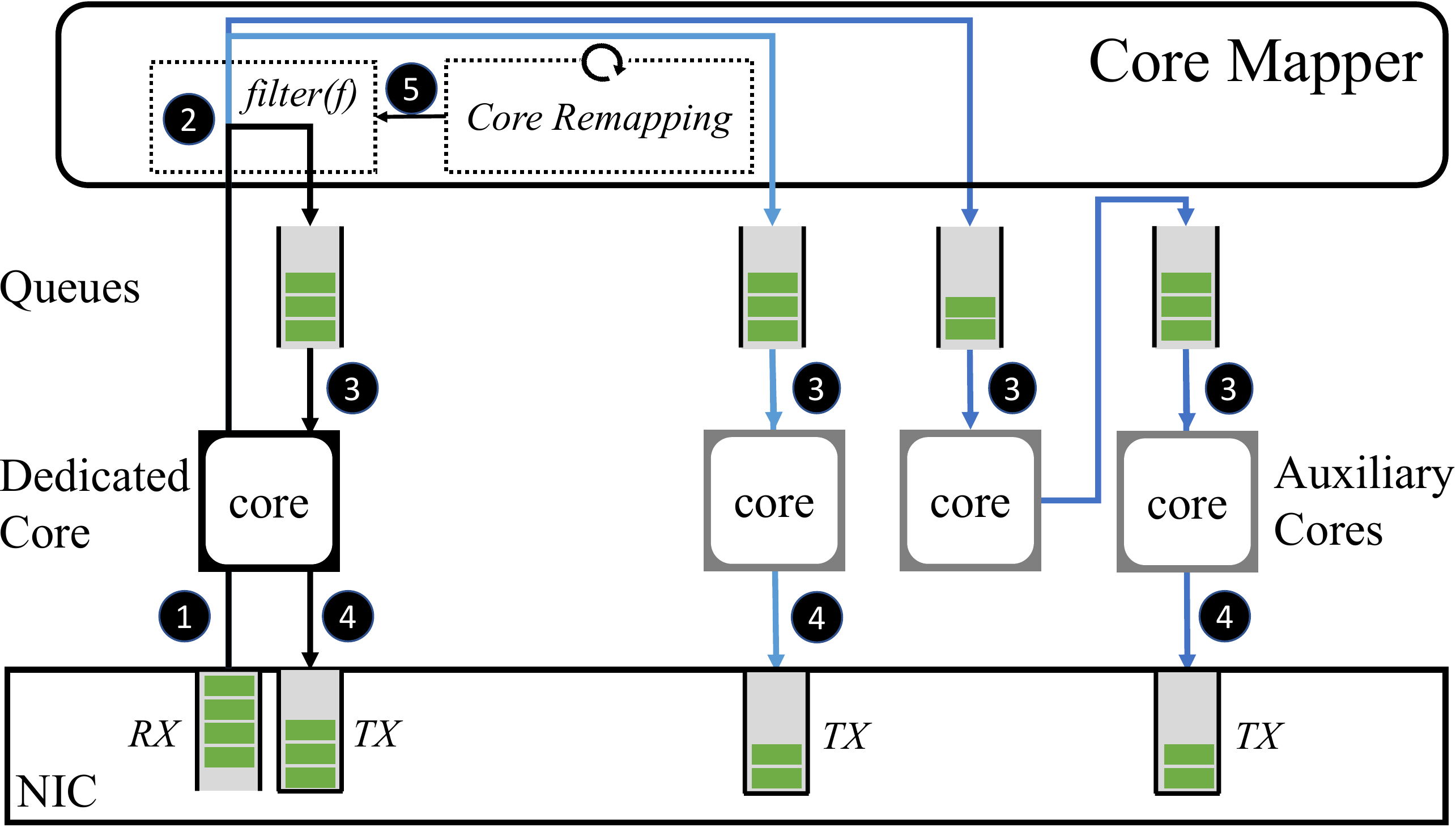}
% }
\caption{For each epoch, the NFV runtime maintains the number of packet and flow arrivals. At the end of an epoch, the core mapper flags potential SLO violations. If the backlog exceeds the core's capacity, the core mapper splits the queue into multiple smaller ones, and recruiting some auxiliary cores to handle them.}
\label{fig:short-term-pipeline}
\end{figure}

\parab{Goal and Challenge.} The server mapper assigns a set of flows mapped to a NIC queue to a dedicated core. The core mapper, one instance of which runs on each dedicated core, seeks to prevent SLO violations on each of its assigned flows. To do this, it must make flow re-mapping decisions at timescales smaller than the target latency SLO (of 100s of $\mu$s). In principle, the core mapper is similar to microsecond-scale core allocation for RPCs~\cite{nsdi19-shenango,osdi20-caladan}, but the flow-affinity constraint results in a qualitatively different design.

\parab{Approach.} The core mapper continuously and at microsecond scales (a) determines potential SLO violations, (b) predicts additional core capacity required to avoid these violations, and (c) re-maps some of its flows to one or more auxiliary cores.

\parab{Determining potential SLO violations.} Given a \pninetynine latency SLO of $L$, \sysname must monitor traffic (and remap flows) at timescales shorter than $L$ in order to detect violations. While many choices are possible for the timescale, a sufficient one is a timescale of $\frac{L}{2}$. This interval is called an \textit{epoch}. For example, if $L$ is 200 $\mu$s, the core mapper uses an epoch of 100 $\mu$s. Within each epoch, the core mapper seeks to maintain the following invariant:
it allocates enough CPU cores to process, within the epoch, any flow that  arrives before the beginning of the epoch. With this invariant, it can ensure that no packet stays in the core for more than two epochs, which is the latency SLO. These epoch durations are 3-4 orders of magnitude smaller than those of other NFV systems~\cite{conext19-rsspp,socc22-quadrant,infocom22-dyssect}.

To determine potential SLO violations, at the end of every epoch, \sysname:
(a) counts the number of packets and active flows in the epoch; and
(b) predicts whether the dedicated core can process these without causing SLO violations.

\parae{Determining queue occupancy.} At each dedicated core, the NFV runtime\footnote{The runtime delivers packets to be processed by the NF chain.} continuously pulls packets from the NIC queue, and tracks packet and flow arrivals (\figref{fig:short-term-pipeline}). It also tracks which packets and which flows have been processed within an epoch. From these, it estimates the count of packets, and the number of new flow arrivals in the queue during the current epoch (the \textit{backlog}).

\parae{Predicting core usage.} Using these two quantities, \sysname determines whether the dedicated core can handle the backlog. If not, it must recruit auxiliary cores. \sysname trains a data-driven predictor obtained from real traces. This training works as follows. \sysname processes the real trace using the complete system (including ingress and server mappers), but disables the core mapper's flow migration component (discussed below). This results in epochs with SLO violations. In each such epoch, it records the number of flows ($f$) and packets ($p$) the core was actually able to process.

\begin{figure}[t]
  \centering
  \resizebox{0.95\columnwidth}{!}{
    \centering
    \pgfplotstableread[row sep=\\,col sep=&]{
p   & sriov64 \\
1   & 9000 \\
10  & 9000 \\
20  & 9100 \\
30  & 9100 \\
40  & 9200 \\
50  & 9200 \\
60  & 9300 \\
70  & 9300 \\
80  & 9400 \\
90  & 9400 \\
95  & 9500 \\
99  & 11400 \\
}\delaymeas

\definecolor{bluegray}{rgb}{0.4, 0.6, 0.8}

\begin{tikzpicture}
    \pgfplotstableread[col sep=&]{figs/data/short_profile_example.dat}\shortprofileone;
  \begin{axis}[
    width = 0.7\textwidth,
    height = 0.35\textwidth,
    Marker/.style = {
      only marks,
      mark = *,
      forget plot,
    },
    EcdfLine/.style = {
      bluegray,
      draw,
    %   ultra thick,
      line width=1.25pt,
    },
    VirtLine/.style = {
      EcdfLine,
      mark = *,
      mark size=1.25pt,
      mark options={fill=green},
    },
    font = \large,
    xtick pos=bottom,
    xmajorgrids,
    ymajorgrids,
    grid style=dashed,
    major tick length = 2,
    minor tick length = 1,
    ytick={0,32,64,...,196},
    minor ytick = {0,16,32,...,196},
    ymin=30,
    ymax=170,
    xmin = 1,
    xmax = 85,
    xtick = {1,10,20,30,...,80},
    minor xtick = {0,5,...,80},
    xtick align=outside,
    ylabel={Number of Packets},
    xlabel={Number of Active Flows},
    legend image post style={scale=1.5},
    legend cell align = left,
    legend style={
      font=\large,
      at={(.565,.86)},
      anchor=west},
    ]
    \addplot[VirtLine] table [x expr=\thisrowno{0}, y expr=\thisrowno{1}] {\shortprofileone};
    \addlegendentry{Epoch Size: 200~$\mu$s};
  \end{axis}
\end{tikzpicture}

%%% Local Variables:
%%% mode: latex
%%% TeX-master: "../main"
%%% End:
  }
  \caption{An example of the Pareto-frontier for the number of active flows (with at least one packet arrival) ($f$) and the number of packets ($p$) that a test NF chain is able to process within an epoch. As $f$ increases, the chain must process less packets in order to avoid backlog.}
\label{fig:short-term-profile-example}
\end{figure}
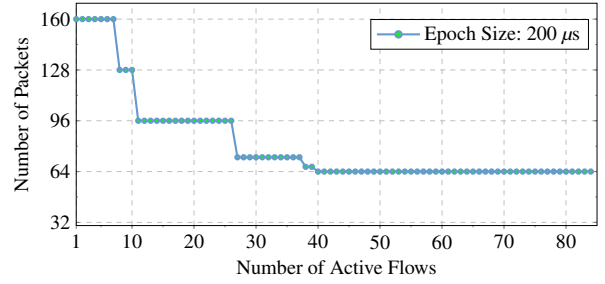

Each such tuple $\tuple{f}{p}$ represents a measure of the \textit{capacity} of the core and the collection of tuples forms a Pareto-frontier. Any $\tuple{f}{p}$ outside this frontier cannot be processed by the core. For example, if a tuple $\tuple{55}{64}$ is obtained during training, the core cannot process a backlog of $\tuple{55}{220}$; it must recruit one or more auxiliary cores to process the \textit{residual backlog}. This intuition forms the basis for the core mapper's capacity predictor, which it uses to determine how much of the backlog can be processed by the dedicated core. It also uses this to determine how many auxiliary cores are needed to process the residual backlog. It does this by repeatedly assigning part of the residual backlog to an auxiliary core, using the predictor, until all the backlog has been assigned. \figref{fig:short-term-profile-example} shows the Pareto-frontier for a stateful NF chain used in \secref{sec:evaluation}) under an epoch size of 200~$\mu$s.\footnote{The Pareto-frontier exhibits discrete jumps of 32 packets, since packets are processed in batch (maximum batch size is 32)}

Thus, at the end of this step, core mapper develops an assignment of $\tuple{f'_{i}}{p'_{i}}$ tuples to the dedicated core and to each auxiliary core (the subscript $i$ denotes the corresponding core), such that the sum of the $f'$s and $p'$s correspond to the backlog. In the above example, the original backlog may be split into three groups of flows: $\tuple{26}{95}$, $\tuple{22}{80}$, and $\tuple{7}{45}$ so that each group of flows is within the Pareto-frontier and can be processed by a core in an epoch.

% \ramesh{Instead of pseudo code, we might consider adding an example. This example can show how, for one dedicated core, some flows are moved to a separate core.}

\parab{Core re-mapping.} Now, \sysname must migrate $f'_{i}$ flows with $p'_{i}$ packets to the $i$-th auxiliary core. To do this, it recruits one (or more) auxiliary cores from a pool. For each auxiliary core, the core mapper \textit{migrates} a subset of flows in its backlog to a \textit{software queue} associated with the auxiliary core. To migrate a flow, the core mapper creates a software queue associated with the auxiliary core, and moves packets from its own queue to that software queue. In turn, each auxiliary core runs a core mapper to avoid SLO violations due to traffic bursts in subsequent epochs.

The tricky step in core re-mapping is handling whales. A whale is a flow whose packet backlog is so high that a single core cannot drain it in one epoch. Since \sysname must maintain flow-affinity, it can only \textit{split} NFs: if a chain has 3 NFs, we can either move the last 1 or 2 NFs to an auxiliary core. This ensures that more packets can be processed, so \sysname can drain the backlog. To do this, given a particular packet backlog $p$ for a whale flow, \sysname needs to determine the best splitting strategy to use. It does this by developing another predictor by profiling different packet counts and different splitting strategies. In the event that the packet count cannot be satisfied by any profiled splitting strategy, \sysname processes the flow best-effort (which can result in SLO violations), but also notifies the operator.

\subsection{Core Efficiency: The Server Mapper}
\label{s:core-effic-serv}

\parab{Goal and Challenge.} The core mapper makes decisions every 100s of $\mu$s and can handle bursts on that timescale, but flow rates and counts may change over longer timescales. The server mapper seeks to ensure high core efficiency by continuously re-allocating flows to the minimum number of dedicated cores necessary to process the offered load.

If we could do this at $\mu$s timescales we would not need the core mapper, but it is non-trivial to re-map large traffic aggregates at those time-scales: software-based forwarding (e.g., via vSwitch) is expensive and hard to scale, especially when a NIC's line rate reaches 100~Gbps. Hardware-based approaches rely on reprogramming flow tables at NICs or switches, which cannot be done at $\mu$s timescales. Thus, we choose a hierarchical design: the server mapper provisions dedicated cores, and the core mapper recruits auxiliary cores to clear short-term backlogs.

The key challenge for the server mapper is to (a) quickly estimate the number of dedicated cores necessary to serve the offered load and (b) rapidly re-compute flow-to-core mappings. The smaller the timescale at which it can do this, the higher \sysname's CPU efficiency. In our current design, the server mapper runs every 1 second. Smaller intervals do not further improve the system's core efficiency and may make \sysname unstable because measurements of packet rate and flow rate can become inaccurate at smaller timescales.

\parab{Dedicated core count predictor.} To predict the number of dedicated cores, \sysname needs a core capacity predictor. The core mapper uses such a predictor, but that predicts the backlog that a core can drain within a short time interval. Unlike the core mapper, the server mapper needs a predictor that can estimate what offered load (combination of active flow count $f$ and aggregate packet rate $r$) a core can serve without being overloaded at a longer timescale. To obtain this, for various values of $f$, for each NF chain and its associated latency SLO, \sysname empirically determines what packet rate $r$ the core can support without violating the latency SLO. Using this data, given an aggregate flow count $f$ and an aggregate packet count $p$, \sysname distributes this offered load across as few dedicated cores as possible; we describe this below.

For this predictor, \sysname uses synthetic traffic, which does not capture the burstiness in real traffic. For this reason, the server mapper's estimate is conservative: it does not aim to guarantee the tail latency target, instead, it relies on the core mapper to adapt to bursty traffic on-demand.
% In fact, we find that the server mapper's predictor does not have to be as accurate as the core mapper's (see experiments in~\secref{subsec:eval-server-mapper}).
% To keep our design simple, the predictor uses the minimal packet rate $r_{min}$ across all possible aggregate flow counts, and does not have to consider various flow counts.

\parab{Fast flow-to-core remapping.} This is the server mapper's primary function. The server mapper knows the aggregate flow count $f'$ and packet rate $r'$ of traffic at each active dedicated core. Its core capacity predictor can determine, for a given core, what flow count $\hat{f}$ and packet rate $\hat{r}$ can be supported on the core. It then seeks to (re)distribute incoming traffic to the fewest number of dedicated cores possible.

% As discussed above, the more frequently the server mapper can redistribute traffic, the better it can track demands and ensure CPU efficiency. 

\parae{Leveraging RSS for fast mapping.} We borrow and adapt an idea from prior work~\cite{conext19-rsspp,infocom22-dyssect}. Many NICs support Receive-Side-Scaling (RSS)~\cite{rss}, which hashes incoming traffic into a large (\eg 128 or 512) number of RSS buckets. Each bucket, in turn, can be bound to a NIC queue assigned to one dedicated core; traffic from this bucket is destined to that core. In general, since the number of buckets is larger than the number of cores, multiple buckets may be bound to a core. \sysname leverages this observation, albeit differently from~\cite{conext19-rsspp}.

\parae{Fast heuristics for traffic re-distribution.}
For each RSS bucket, \sysname tracks the average active flow count and packet rate for each decision interval. Then, for each core, it needs to find a group of buckets satisfying the following constraints:
\begin{itemize}[nosep,leftmargin=*]
\item The group of buckets is uniquely assigned to the core.
\item The sum of each bucket's $\tuple{f}{r}$ tuple, denoted as $\tuple{f'}{r'}$, does not exceed the predicted capacity $\tuple{\hat{f}}{\hat{r}}$.
\end{itemize}

Besides these, the server mapper has to ensure that all buckets are assigned to cores. It is possible to model this problem as an optimization formulation. The objective is to minimize the number of dedicated cores. Unfortunately, this formulation (which we omit for brevity) is a constrained mixed-integer linear problem (MILP), which can take hundreds of ms or more to solve using a commercial solver.

Instead, we use a fast greedy heuristic, which attempts to use the fewest dedicated cores, but models a key constraint that is hard to express in a mathematical formulation: minimizing the number of changes to bucket-to-core mappings. This ensures that flows do not need to be frequently migrated between cores, which can degrade the performance of NF processing as it impacts cache locality.

The heuristic handles cores whose load has increased (because of flow arrivals or traffic increases), as well as those whose load has decreased (because of departures):
\begin{itemize}[nosep,leftmargin=*]
\item It first determines which dedicated cores are \textit{overloaded} (\ie their $\tuple{f'}{r'}$ exceeds core capacity). For each, it determines the fewest buckets that need to be migrated away to other dedicated cores to ensure the remaining traffic fits within the core capacity. It moves these buckets to other under-loaded cores using a first-fit strategy. If no such core exists, it allocates a new dedicated core.
\item Then, it starts with the least loaded core and attempts to migrate its existing buckets to other cores, reclaiming the core if successful. It repeats this until no cores can be reclaimed.
\end{itemize}

% \begin{figure}[tb!]
% \resizebox{\columnwidth}{!}{
%     \includegraphics[width=\textwidth]{figs/ironside-rss.pdf}
% }
% \caption{The server mapper updates the NIC's indirection table to apply bucket-to-core mappings. For each decision interval, it tracks the flow count and packet rate $<f,r>$ for each RSS bucket, corresponding to one entry in the NIC's table. At the end of an interval, the server mapper decides the number of dedicated cores at a server: it finds overloaded cores, whose loads exceed the core's capacity, moves the minimum set of buckets from them to other cores, and then tries to reclaim cores. In the above case, one bucket is migrated from core 1 to core 2 to avoid overloading.}
% \label{fig:server-mapper}
% \end{figure}

\sysname writes the new bucket-to-core mapping table before deleting the old one to ensure no packets are dropped. This step, because of NIC limitations, can take up to a few milliseconds~\cite{conext19-rsspp,pam21-nicbench}, . It conservatively chooses a decision interval of 1~s, to minimize overhead, and to avoid oscillations in bucket-to-core assignments. \figref{fig:server-mapper} in the appendix illustrates this algorithm.

\parab{Boost mode for dedicated cores.} In between periodic invocations of the server mapper, a dedicated core can see large backlogs that last for more than a few epochs. \sysname must handle such overloads to minimize SLO violations, since  a dedicated core can dispatch only a limited number of packets within a epoch.

To address this, we tried to invoke the server mapper on-demand (in between two periodic invocations). However, this still led to a high tail latency because of the millisecond-level delay of updating the RSS table in NIC hardware, as we mentioned above.
To avoid this, under such overload, \sysname uses a \textit{boost mode} for dedicated cores in software. In this mode, the dedicated core skips NF processing and focuses on pulling packets from the NIC queue and dispatching them to auxiliary cores. To do this, it recruits another core to conduct NF processing on its behalf. Once the queue has been drained, it can resume NF processing. In evaluation, we show that enabling boost-mode is necessary for meeting tail latency SLOs (\secref{subsec:eval-server-mapper}).

% Persistent load changes can happen in between two consecutive long-term optimizations, because of significantly increased packet rates of existing flows and an increased flow arrival rate.
% In such cases, one normal core may observe a large software queue, which then requires \sysname's short-term optimization to migrate a subset of flows from the queue.
% However, the short-term optimization can introduce large overheads when many packets present in a software queue, as the short-term optimization expects rare bursts in traffic and assumes small software queues, and fails to absorb the large queue effectively within a small number of short-term epochs.
% Therefore, when a normal core observes a large queue (e.g., 256 packets in a normal core's software queue), it will trigger a \sysname's long-term optimization immediately\footnote{\sysname has to respect the hardware limitation of modifying rules at the NIC's flow table. At each server, we set a lower bound on the time interval of two long-term optimizations as 5 milliseconds (for Mellanox ConnectX-5 100 GbE).} to split its traffic into two subsets. One subset is marked to be migrated by the above two-step heuristics. After that, the current long-term epoch is reset.

\subsection{Minimizing Servers: The Ingress Mapper}
\label{s:ingress-mapper}
% why are we doing this:
% two different granularity of traffic assignment?

% what is the system deployment scenario

The ingress mapper assigns flows to servers with the goal of minimizing the number of servers assigned to the NF chain. A rack may see a large number of flows, so the ingress mapper cannot scale if it needs to make per-flow server assignment decisions. Instead, the ingress mapper makes assignment decisions on a prefix of the destination address. Specifically, it assigns flows with the same destination prefix of length $k$ to the same server. This requires $2^{32-k}$ routing entries for IPv4. $k$ can be chosen to match ToR switch table capacities.

When traffic arrives at a previously \textit{unoccupied} (\ie one for which no packets have been seen) prefix, it steers that prefix to the server with the most number of dedicated cores as long as that number is under a threshold $\tau$. If no such server exists, the ingress mapper finds a new one from the rack and steers traffic to it. Thus, new servers are recruited only when all other servers are near the target load. Parameter $\tau$ ensures that each server has sufficient auxiliary cores to handle bursts.

When it changes the number of dedicated cores, each server mapper reports this number to the ingress mapper, which uses this to steer prefixes. \sysname does not attempt to consolidate servers with few dedicated cores. This requires cross-server NF state migration, but is also unnecessary, since we assume that non-dedicated cores (on any server) can be used to run edge applications.

\section{Evaluation}
\label{sec:evaluation}

In this section, we evaluate:
 \sysname's ability to achieve microsecond-scale tail latency SLOs  (\secref{subsec:eval-scaling-latency}),
\sysname's CPU core efficiency relative to other NFV systems (\secref{subsec:eval-scaling-core-usage}), and how
its design choices contribute to its performance (\secref{subsec:eval-core-mapper}-\secref{subsec:eval-server-mapper}).

\parab{Implementation.} \sysname uses BESS~\cite{bess-git}, a high-performance DPDK-based switch~\cite{dpdk}. 
Its design consists of two parts: a rack-scale ingress and a \textit{worker}, requiring 2.9k lines of C++. The former implements statistics collection from workers and runs the ingress mapper (\secref{s:ingress-mapper}). The latter contains the server mapper (\secref{s:core-effic-serv}), the core mapper (\secref{s:absorb-bursts:-core}), and additional functionality to enable dedicated cores to pull traffic from the NIC queues into software queues, as well as to recruit auxiliary cores to handle bursts. In addition, \sysname's  framework for training core capacity predictors adds 1.2k lines of Python.

\subsection{Setup and Methodology}
\label{subsec:eval-methodology}

% r6525 32 nodes (AMD EPYC Milan, 64 core, 256GB RAM, 1.6TB NVMe)
\parab{Testbed.} We use Cloudlab~\cite{cloudlab-paper} to run experiments on a cluster of 5 servers under a single ToR. Each server has dual-CPU 32-core 2.8 GHz AMD EPYC Milan CPUs with 256~GB ECC memory (DDR4 3200~MHz). To reduce jitter, we disable SMT and CPU frequency scaling. Each server has a 100~GbE Mellanox ConnectX-6 NIC for data traffic and a separate NIC for control and management traffic. %Each machine has one Mellanox ConnectX-5 NIC, connected to a separated LAN for control and management traffic.

\parab{Methodology.} On this cluster, we run \sysname and other NFV systems to serve NF chains that process real-world traffic, and measure end-to-end packet processing latency and CPU core usage metrics. We use two canonical stateful NF chains, from documented use cases~\cite{ietf-sfc-use-cases-06,nsdi20-batchy}.

\textit{\textbf{Chain 1}} is a flow monitoring and load-balancing pipeline: ACL$\rightarrow$LB$\rightarrow$Accounting. \textit{\textbf{Chain 2}} is a more compute-intensive chain with DPI and encryption NFs: ACL$\rightarrow$UrlFilter$\rightarrow$Encrypt. These NFs and NF chains are frequently used for evaluating NFV systems' performance~\cite{nsdi18-metron,socc20-snf,socc22-quadrant}.
ACL enforces 1.5k access control rules. LB is a L4 load balancer that distributes flows across a list of backend servers. Accounting is a monitoring NF that tracks the data usage of each flow. UrlFilter performs TCP reconstruction over flows, and applies complex string matching rules (\ie Snort [45] rules) to block connections mentioning banned URLs. Encrypt encrypts each packet with 128-bit ChaCha~\cite{bernstein2008chacha}. NAT maintains a list of available L4 ports and performs address translation for connections, assigning a free port and maintaining this port mapping for a connection’s lifetime. 

\parab{Traffic traces.}  Our experiments use two real traffic traces\footnote{While we target \sysname for edge settings, in the evaluation we use backbone and AS traces since edge traffic in a few years may increase to those levels.}  (\secref{sec:motivation}). The \textit{\textbf{backbone trace}} contains 4.9M flows, with flow sizes from 54 bytes to 206M bytes, and a maximum duration of 52 seconds. The flow arrival rates are 31.2k (average) and 51.2k (max) flows per second, and packet sizes are 988 (average) and 1514 (max) bytes. The \textit{\textbf{AS trace}} contains 1.6M flows ranging from 54 bytes to 514M bytes in size, with a maximum duration of 205 seconds. The flow arrival rates are 10.1k (average) and 39.0k (max) per second. The packet sizes are 976 (average) and 1514 (max) bytes. Thus, the AS trace exhibits a larger range of flow sizes and durations, but slightly lower flow rates, than the backbone trace. We train our core capacity predictors (\secref{s:absorb-bursts:-core}, \secref{s:core-effic-serv}) on 15~s of traffic the backbone trace, and test on the rest.

We run a BESS-based traffic generator on a dedicated server. Each experiment runs a single combination of NF chain and traffic trace. For our choice of chains and traffic traces, 5 servers suffice to handle the workload, hence our choice of cluster size.
% To produce a large traffic volume, we enable the multi-core acceleration for , and run it on a dedicated machine.

\begin{table}[t!]
  \small
\centering
\resizebox{0.98\columnwidth}{!}{
\begin{tabular}{|m{1.45cm}|m{1.25cm}|*{4}{>{\centering\arraybackslash}m{1.0cm}}|}
  \hline
  \textbf{Platform} & \centering \textbf{Latency SLO} & \textbf{p50} & \textbf{p99} & \textbf{Avg. Cores} & \textbf{Loss Rate} \\
   & \centering{[$\mu$s]} & \centering{[$\mu$s]} & \centering{[$\mu$s]} & & \\
  \hline
  \hline
  \multirow{4}{*}{\textbf{{\sysname}}} & \textit{100} & 28.8 & 89.4 & 34.6 & 0.006 \\
  \cline{2-6} & \textit{200} & 32.1 & 174.4 & 30.7 & 0.006 \\
  \cline{2-6} & \textit{300} & 30.7 & 238.8 & 30.1 & 0.005  \\
  \cline{2-6} & \textit{400} & 34.5 & 254.4 & 30.5 & 0.010  \\
  \cline{2-6} & \textit{500} & 31.1 & 298.2 & 29.7 & 0.005  \\
  \hline
  \hline
  \textbf{{Metron}} & - & 57.7 & 2541 & 21.8 & 0.024 \\
  \hline
  \hline
  \multirow{4}{*}{\textbf{{Quadrant}}} & \textit{100} & 30.8 & 3051 & 45.3 & 0.062 \\
  \cline{2-6} & \textit{200} & 39.1 & 3021 & 43.3 & 0.057  \\
  \cline{2-6} & \textit{300} & 38.8 & 3015 & 43.3 & 0.056  \\
  \cline{2-6} & \textit{400} & 39.9 & 2988 & 43.1 & 0.052  \\
  \cline{2-6} & \textit{500} & 53.7 & 2994 & 41.7 & 0.044  \\
  \hline
  \hline
  \multirow{4}{*}{\textbf{{Dyssect}}} & \textit{100} & 63.0 & 78798 & 34.7 & 0.224 \\
  \cline{2-6} & \textit{200} & 68.0 & 68356 & 31.9 & 0.106  \\
  \cline{2-6} & \textit{300} & 78.1 & 70392 & 32.7 & 0.249  \\
  \cline{2-6} & \textit{400} & 76.2 & 64885 & 30.1 & 0.218  \\
  \cline{2-6} & \textit{500} & 68.7 & 71586 & 31.3 & 0.194  \\
  \hline
\end{tabular}
}
\caption{Comparisons of end-to-end metrics (p50 and p99 latency, the time-averaged CPU core usage and loss rate) for \textbf{chain 1} under the \textbf{backbone traffic} by \sysname and others, as a function of the system's latency target.}
\label{t:latency-chain1-backbone-traffic}
\vspace{1mm}
\end{table}

\parab{Metrics.} We quantify \sysname performance using: (a) \textit{\textbf{end-to-end latency}}, for which we use two measures: the median (\pfifty) and the \pninetynine tail latency; and (b) \textit{\textbf{CPU core usage}} in terms of total CPU-hour(s) or the time-averaged CPU cores for serving test traffic. Ideally, \sysname should exhibit low CPU core usage while ensuring the target \pninetynine latency SLO.

\parab{Comparisons.} We compare \sysname against three prior NFV systems that auto-scale NF chains under traffic dynamics.

\textit{\textbf{Metron}}~\cite{nsdi18-metron} supports scaling of NF chains. It enables L2 switching at NICs, and modifies L2 headers at the ToR switch to enforce routing decisions. It detects overloaded cores by periodically checking packet losses (via packet counters) at the switch, and splits a core's traffic aggregate into two subsets and migrates one subset to a new core when it detects core overload. Metron does not explicitly support latency SLOs.

\textit{\textbf{Quadrant}}~\cite{socc22-quadrant} explicitly targets \pninetynine latency SLOs, scales NF chains by collecting the worst-case packet delay at each core via its cluster-scale ingress, and uses a hysteresis-based approach to control end-to-end latency of each core running a chain. It has been shown to satisfy \pninetynine latency SLOs for synthetic traffic. Real traffic traces contain significant bursts that can, as we show below, impact Quadrant's performance.

\textit{\textbf{Dyssect}}~\cite{infocom22-dyssect} seeks to minimize \pfifty latency of serving NF chains. It uses RSS to scale NF chains at a worker, and periodically migrates a subset of flows to new CPU cores, using an MILP formulation. However, Dyssect is not designed for a rack-scale system; to enable the rack-scale experiments for Dyssect, we apply \sysname's rack-scale ingress to distribute traffic to Dyssect workers.

% what graphs to produce?
% inputs: system, NF chain, traffic, SLO
% 1. latency CDF for a system, a chain, a traffic, a slo
% 2. CPU core usage for a system, a chain, a traffic, a slo

\begin{table}[t!]
\small
\centering
\resizebox{0.98\columnwidth}{!}{
\begin{tabular}{|m{1.45cm}|m{1.25cm}|*{4}{>{\centering\arraybackslash}m{1.0cm}}|}
  \hline
  \textbf{Platform} & \centering \textbf{Latency SLO} & \textbf{p50} & \textbf{p99} & \textbf{Avg. Cores} & \textbf{Loss Rate} \\
   & \centering{[$\mu$s]} & \centering{[$\mu$s]} & \centering{[$\mu$s]} & & \\
  \hline
  \hline
  \multirow{4}{*}{\textbf{{\sysname}}} & \textit{100} & 18.0 & 101.4 & 56.0 & 0.004 \\
  \cline{2-6} & \textit{200} & 22.8 & 180.3 & 41.9 & 0.010 \\
  \cline{2-6} & \textit{300} & 24.8 & 197.2 & 42.8 & 0.002  \\
  \cline{2-6} & \textit{400} & 24.9 & 210.3 & 40.3 & 0.007  \\
  \cline{2-6} & \textit{500} & 26.8 & 270.5 & 40.6 & 0.003  \\
  \hline
  \hline
  \textbf{{Metron}} & - & 254.5 & 5580 & 32.1 & 0.044 \\
  \hline
  \hline
  \multirow{4}{*}{\textbf{{Quadrant}}} & \textit{100} & 34.0 & 5459 & 76.7 & 0.126 \\
  \cline{2-6} & \textit{200} & 34.7 & 5465 & 78.3 & 0.133  \\
  \cline{2-6} & \textit{300} & 34.0 & 5452 & 75.5 & 0.132  \\
  \cline{2-6} & \textit{400} & 34.5 & 5473 & 77.1 & 0.137  \\
  \cline{2-6} & \textit{500} & 34.2 & 5455 & 77.2 & 0.135  \\
  \hline
  \hline
  \multirow{4}{*}{\textbf{{Dyssect}}} & \textit{100} & 62.6 & 78955 & 50.6 & 0.294 \\
  \cline{2-6} & \textit{200} & 61.0 & 62528 & 45.6 & 0.246  \\
  \cline{2-6} & \textit{300} & 60.2 & 77793 & 47.4 & 0.279  \\
  \cline{2-6} & \textit{400} & 50.2 & 87130 & 48.2 & 0.208  \\
  \cline{2-6} & \textit{500} & 61.4 & 78030 & 44.8 & 0.214  \\
  \hline
\end{tabular}
}
\caption{Comparisons of end-to-end metrics (p50 and p99 latency, the time-averaged CPU core usage and loss rate) for \textbf{chain 2} under the \textbf{backbone traffic} by \sysname and others, as a function of the system's latency target.}
\label{t:latency-chain2-backbone-traffic}
\vspace{1mm}
\end{table}

\subsection{Latency Performance Comparisons}
\label{subsec:eval-scaling-latency}

% \begin{figure*}[!th]
% \centering
% \resizebox{\textwidth}{!}{\input{figs/latency_traffic1.tex}}
% \caption{Comparisons of end-to-end latency achieved by \sysname and other NFV platforms, as a function of latency SLO.}
% \label{fig:latency_results}
% \end{figure*}

We first compare \sysname's latency against other NFV systems that use different auto-scaling mechanisms. 

\parae{Results.} Tables~\ref{t:latency-chain1-backbone-traffic},~\ref{t:latency-chain2-backbone-traffic},~\ref{t:latency-chain1-as-traffic},and~\ref{t:latency-chain2-as-traffic} show the p50 and p99 latency of different NFV platforms when serving the backbone traffic and the AS traffic. Across all platforms, \sysname is the only one capable of achieving microsecond-scale (100-500 $\mu$s) p99 latency SLOs. All platforms
do exhibit sub-millisecond p50 latency, showing that their scaling techniques work well to bound median latency. \sysname achieves this while also incurring the lowest packet loss rate among all systems (see \textbf{Loss Rate} column).

% traffic 2: chain 1
\begin{table}[t]
  \small
\centering
\resizebox{0.98\columnwidth}{!}{
\begin{tabular}{|m{1.45cm}|m{1.25cm}|*{4}{>{\centering\arraybackslash}m{1.0cm}}|}
  \hline
  \textbf{Platform} & \centering \textbf{Latency SLO} & \textbf{p50} & \textbf{p99} & \textbf{Avg. Cores} & \textbf{Loss Rate} \\
   & \centering{[$\mu$s]} & \centering{[$\mu$s]} & \centering{[$\mu$s]} & & \\
  \hline
  \hline
  \multirow{4}{*}{\textbf{{\sysname}}} & \textit{100} & 22.7 & 51.9 & 17.5 & 0.010 \\
  \cline{2-6} & \textit{200} & 22.9 & 52.1 & 17.4 & 0.011 \\
  \cline{2-6} & \textit{300} & 23.1 & 53.4 & 15.0 & 0.006  \\
  \cline{2-6} & \textit{400} & 23.4 & 53.2 & 16.2 & 0.013  \\
  \cline{2-6} & \textit{500} & 23.5 & 53.5 & 17.5 & 0.008  \\
  \hline
  \hline
  \textbf{{Metron}} & - & 48.6 & 1926 & 11.9 & 0.014 \\
  \hline
  \hline
  \multirow{4}{*}{\textbf{{Quadrant}}} & \textit{100} & 37.2 & 1026.5 & 17.9 & 0.014 \\
  \cline{2-6} & \textit{200} & 30.7 & 930.0 & 17.3 & 0.010  \\
  \cline{2-6} & \textit{300} & 29.5 & 765.2 & 16.9 & 0.013  \\
  \cline{2-6} & \textit{400} & 29.3 & 556.5 & 17.3 & 0.011  \\
  \cline{2-6} & \textit{500} & 28.2 & 810.2 & 16.4 & 0.011  \\
  \hline
  \hline
  \multirow{4}{*}{\textbf{{Dyssect}}} & \textit{100} & 17.7 & 13586 & 17.9 & 0.044 \\
  \cline{2-6} & \textit{200} & 17.6 & 13031 & 17.2 & 0.054  \\
  \cline{2-6} & \textit{300} & 17.8 & 12672 & 17.9 & 0.050  \\
  \cline{2-6} & \textit{400} & 18.2 & 12480 & 15.9 & 0.052  \\
  \cline{2-6} & \textit{500} & 19.1 & 22776 & 16.5 & 0.049  \\
  \hline
\end{tabular}
}
\caption{Comparisons of end-to-end metrics (p50 and p99 latency, the time-averaged CPU core usage and loss rate) for running \textbf{chain 1} under the \textbf{AS traffic} by \sysname and others, as a function of the system's latency target.}
\label{t:latency-chain1-as-traffic}
\end{table}

% traffic 2: chain 2
\begin{table}[t]
\small
\centering
\resizebox{0.98\columnwidth}{!}{
\begin{tabular}{|m{1.45cm}|m{1.25cm}|*{4}{>{\centering\arraybackslash}m{1.0cm}}|}
  \hline
  \textbf{Platform} & \centering \textbf{Latency SLO} & \textbf{p50} & \textbf{p99} & \textbf{Avg. Cores} & \textbf{Loss Rate} \\
   & \centering{[$\mu$s]} & \centering{[$\mu$s]} & \centering{[$\mu$s]} & & \\
  \hline
  \hline
  \multirow{4}{*}{\textbf{{\sysname}}} & \textit{100} & 33.1 & 87.0 & 25.2 & 0.006 \\
  \cline{2-6} & \textit{200} & 41.6 & 103.3 & 18.1 & 0.008 \\
  \cline{2-6} & \textit{300} & 48.8 & 105.6 & 16.3 & 0.008  \\
  \cline{2-6} & \textit{400} & 50.0 & 108.0 & 18.4 & 0.007  \\
  \cline{2-6} & \textit{500} & 60.8 & 103.3 & 16.6 & 0.007  \\
  \hline
  \hline
  \textbf{{Metron}} & - & 78.7 & 5310 & 14.8 & 0.029 \\
  \hline
  \hline
  \multirow{4}{*}{\textbf{{Quadrant}}} & \textit{100} & 70.4 & 5352 & 19.2 & 0.073 \\
  \cline{2-6} & \textit{200} & 69.9 & 5363 & 18.6 & 0.072  \\
  \cline{2-6} & \textit{300} & 70.5 & 5351 & 19.3 & 0.074  \\
  \cline{2-6} & \textit{400} & 70.4 & 5369 & 18.2 & 0.073  \\
  \cline{2-6} & \textit{500} & 69.9 & 5373 & 18.2 & 0.072  \\
  \hline
  \hline
  \multirow{4}{*}{\textbf{{Dyssect}}} & \textit{100} & 24.2 & 48041 & 18.4 & 0.056 \\
  \cline{2-6} & \textit{200} & 24.1 & 57273 & 18.1 & 0.049  \\
  \cline{2-6} & \textit{300} & 23.2 & 34573 & 16.9 & 0.041  \\
  \cline{2-6} & \textit{400} & 24.4 & 50812 & 17.0 & 0.040  \\
  \cline{2-6} & \textit{500} & 23.8 & 47229 & 16.7 & 0.044  \\
  \hline
\end{tabular}
}
\caption{Comparisons of end-to-end metrics (p50 and p99 latency, the time-averaged CPU core usage and loss rate) for running \textbf{chain 2} under the \textbf{AS traffic} by \sysname and others, as a function of the system's latency target.}
\label{t:latency-chain2-as-traffic}
\end{table}

Metron, which does not explicitly support SLOs (so has only a single entry in the tables), achieves 2541 $\mu$s and 5580 $\mu$s \pninetynine latency for chain 1 and chain 2 respectively (backbone traffic), and 1926 $\mu$s and 5310 $\mu$s respectively (AS traffic). It also has the highest p50 latency among all platforms because it relies on hardware to achieve dynamic scaling. Detecting packet losses by sending periodic control-plane queries to collect packet counter statistics at ToR switches cannot avoid high latency caused by short-term bursts. Moreover, re-programming switch tables (for splitting traffic aggregates) can take hundreds of milliseconds. 

Quadrant supports p99 latency SLOs targets. However, in our experiments it incurs 2988-3051 $\mu$s (5452-5473 $\mu$s) \pninetynine latency for chain 1 (resp. chain 2) under the backbone traffic. Even though it actively monitors each core's worst-case packet processing delay, it still fails to deliver good tail latency results on highly bursty real-world traces for three reasons. 1) \textit{Improper scaling signal}: for each core, Quadrant maintains a moving average worst-case packet processing delay, updated to a rack-scale ingress every hundreds of milliseconds. This signal is insufficient to detect transient bursts quickly. 2) \textit{Delayed scaling}: Quadrant makes scaling decisions via one rack-scale controller (rather than making local decisions at each worker), which re-balances flows from overloaded cores every 50 milliseconds and enforces scaling via RPCs that add extra millisecond-level delays. 3) \textit{Inaccurate scaling}: Quadrant migrates at-most half of flows from a core in a scaling period, which may not work well because a significant number of flows can arrive within its scaling period. %\sysname, on the other hand, makes local scaling decisions at workers at much smaller timescales, and scales accurately by considering flow and packet counts and using core capacity predictors.

Dyssect supports p50 latency SLO targets. With a goal of optimizing \pfifty latency, it migrates flows from cores to control each core's load, defined as the percentage of CPU time executing actual NF processing. Dyssect produces good \pfifty latency results: 35 $\mu$s and 60 $\mu$s (backbone traffic), but has the highest \pninetynine latency among all: 78 ms and 87 ms for chain 1 and chain 2 (backbone traffic). Core load is an insufficient scaling signal for maintaining low tail latency. Dyssect also uses a 100-ms scaling period, a compute-intensive MILP formulation to determine flow migrations, and locks in its data-plane implementation, all of which increase latency.

In general, for all systems, p99 latencies are higher for chain 2 than chain 1, since the former is more compute intensive. The one exception to this Dyssect whose p99 is comparable for both chains for the backbone trace, which has a much higher flow rate.

% as traffic

To conclude, we find that prior systems are unable to ensure sub-ms tail latency SLOs if (1) they only use hardware-based scaling; (2) they do not have mechanisms to quickly detect and react to traffic bursts. \sysname uses a hybrid scaling design and having the core mapper to process transient bursts quickly.

\subsection{CPU Core Usage Comparisons}
\label{subsec:eval-scaling-core-usage}

To compare CPU core usage of different NFV systems, in each 100-$\mu$s epoch we record the number of active CPU cores in an NFV system, and calculate the time-averaged core usage across the entire experiment. 

\parae{Results.} Tables~\ref{t:latency-chain1-backbone-traffic},~\ref{t:latency-chain2-backbone-traffic},~\ref{t:latency-chain1-as-traffic},~\ref{t:latency-chain2-as-traffic} show CPU core usage for \sysname and other NFV systems.
In addition to having 1-2 orders of magnitude lower tail latency than other NFV systems, \sysname  uses less CPU up to 34\% (chain 1) / 48\% (chain 2) (backbone traffic) and 17\% (chain 1) / 15\% (chain 2) (AS traffic) compared to Quadrant, and uses less CPU up to 14\% (chain 1) and 20\% (chain 2) (backbone traffic) and 16\% (chain 1) / 11\% (chain 2) (AS traffic) compared to Dyssect.  Quadrant monitors per-core \textit{worst-case} packet delay to make flow migration decisions. Transient bursts can increase worst-case delays and trigger flow re-allocation to a new core, which increases core usage. Dyssect's CPU core usage is slightly better than Quadrant's core usage, and comparable to \sysname's. This is because Dyssect uses the CPU core load as its scaling signal, which is less sensitive to bursts. 

Compared to Metron, \sysname uses slightly more (10-26\%) CPU cores (but has an order of magnitude lower \pninetynine latency), as Metron uses efficient hardware-based scaling. Though Metron aims to minimize packet losses during scaling, it does not aggressively optimize end-to-end latency (thus producing higher p50 and p99 latency). In contrast, \sysname achieves an order of magnitude lower p99 latency, prevents packet losses by allocating traffic to cores with core capacity predictors, and demonstrates lower packet loss rates.

To better compare Metron and \sysname core usage, we set \sysname's target p99 latency SLO to be the p99 latency achieved by Metron. \tabref{t:ironside-high-latency-cost} shows that \sysname can still produce lower p99 latency results with loose latency SLOs, and uses a comparable number of CPU cores compared to Metron. In this experiment, \sysname is less aggressive in migrating flows to auxiliary cores, and thus uses fewer cores.

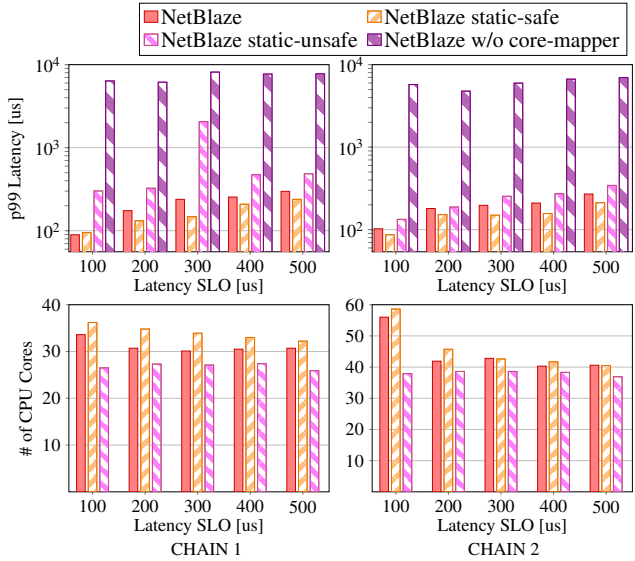
\begin{figure}[t]
    \centering
    \resizebox{\columnwidth}{!}{% traffic 1 / chain 1
\pgfplotstableread[row sep=\\,col sep=&]{
SLO & Loss  & Core & p50  & p99    \\
100 & 0.006 & 33.6 & 28.8 & 89.4   \\
200 & 0.006 & 30.7 & 32.1 & 174.4  \\
300 & 0.005 & 30.1 & 30.7 & 238.8  \\
400 & 0.010 & 30.5 & 34.5 & 254.4  \\
500 & 0.005 & 30.7 & 31.1 & 298.2  \\
}\IronsideTrafficOneChainOne

\pgfplotstableread[row sep=\\,col sep=&]{
SLO & Loss  & Core & p50  & p99      \\
100 & 0.038 & 36.2 & 28.2 & 95.2    \\
200 & 0.038 & 34.8 & 29.6 & 131.6   \\
300 & 0.059 & 33.9 & 30.5 & 147.8   \\
400 & 0.055 & 33.0 & 31.2 & 208.6   \\
500 & 0.053 & 32.2 & 31.6 & 238.6   \\
}\IronsideStaticSafeCoreTrafficOneChainOne

\pgfplotstableread[row sep=\\,col sep=&]{
SLO & Loss  & Core & p50  & p99      \\
100 & 0.044 & 26.5 & 29.3 & 301.3   \\
200 & 0.047 & 27.3 & 30.6 & 324.9   \\
300 & 0.076 & 27.1 & 32.6 & 2056.4   \\
400 & 0.097 & 27.4 & 34.3 & 473.3   \\
500 & 0.085 & 25.9 & 34.3 & 484.9   \\
}\IronsideStaticUnsafeCoreTrafficOneChainOne

\pgfplotstableread[row sep=\\,col sep=&]{
SLO & Loss  & Core & p50  & p99      \\
100 & 0.129 & 16.0 & 31.6 & 6360     \\
200 & 0.137 & 13.7 & 31.3 & 6149     \\
300 & 0.205 & 15.2 & 33.3 & 8137     \\
400 & 0.197 & 15.4 & 34.6 & 7702     \\
500 & 0.197 & 15.2 & 34.6 & 7743     \\
}\IronsideNoCoreTrafficOneChainOne

% traffic 1 / chain 2
\pgfplotstableread[row sep=\\,col sep=&]{
SLO & Loss  & Core & p50  & p99     \\
100 & 0.004 & 56.0 & 18.0 & 102.4   \\
200 & 0.015 & 41.9 & 22.8 & 180.3   \\
300 & 0.006 & 42.8 & 24.8 & 197.2   \\
400 & 0.007 & 40.3 & 24.9 & 210.3   \\
500 & 0.007 & 40.6 & 26.8 & 270.5   \\
}\IronsideTrafficOneChainTwo

\pgfplotstableread[row sep=\\,col sep=&]{
SLO & Loss  & Core & p50  & p99      \\
100 & 0.013 & 58.6 & 18.1 & 86.7   \\
200 & 0.006 & 45.7 & 22.1 & 152.1   \\
300 & 0.015 & 42.6 & 23.3 & 149.7   \\
400 & 0.003 & 41.7 & 23.4 & 156.7   \\
500 & 0.010 & 40.5 & 25.4 & 212.3   \\
}\IronsideStaticSafeCoreTrafficOneChainTwo

\pgfplotstableread[row sep=\\,col sep=&]{
SLO & Loss  & Core & p50  & p99      \\
100 & 0.062 & 37.9 & 21.5 & 133.3   \\
200 & 0.057 & 38.6 & 24.3 & 188.1   \\
300 & 0.056 & 38.6 & 25.8 & 253.8   \\
400 & 0.052 & 38.3 & 26.2 & 272.7   \\
500 & 0.044 & 36.9 & 27.4 & 343.8   \\
}\IronsideStaticUnsafeCoreTrafficOneChainTwo

\pgfplotstableread[row sep=\\,col sep=&]{
SLO & Loss  & Core & p50   & p99      \\
100 & 0.015 & 23.9 & 68.4  & 5738   \\
200 & 0.035 & 21.8 & 106.2 & 4773   \\
300 & 0.086 & 21.1 & 211.6 & 5974   \\
400 & 0.129 & 20.9 & 994.2 & 6670   \\
500 & 0.153 & 21.1 & 1861  & 6963   \\
}\IronsideNoCoreTrafficOneChainTwo

\pgfplotsset{
  % single bar in legend
  /pgfplots/ybar legend/.style={
    /pgfplots/legend image code/.code={
      \draw[##1,/tikz/.cd, yshift=-0.3em]
      (7 mm, 0 mm) rectangle (11pt, 0.75em);},
  },
  CPUUtilGroupPlot/.style = {
    group/group name = cpuutils,
    group/group size = 2 by 2,
    group/xlabels at = edge bottom,
    group/ylabels at = edge left,
    group/vertical sep = 35pt,
  },
  LatencyPlot/.style = {
    font = \large,
    height = .7\columnwidth,
    width = .9\columnwidth,
    bar width=.2cm,
    tick pos = left,
    % xmajorgrids=true,
    ymajorgrids=true,
    major tick length = 2,
    xlabel near ticks,
    ylabel near ticks,
    xtick align = outside,
    ytick align = outside,
    xlabel shift = -3 pt,
    legend columns = 2,
    legend cell align = left,
    legend style = {
       font = \normalsize,
       fill=none,
       at = {(0.98, 1.33), anchor = south},
      /tikz/every even column/.append style={column sep=0.15cm},
    }
  },
  CPUUtilPlot/.style = {
    font = \large,
    legend columns = 4,
    legend cell align = left,
    legend style = {
       font = \footnotesize,
       at = {(0.98, 1.3), anchor = south},
      /tikz/every even column/.append style={column sep=0.10cm},
    },
    height = .5\columnwidth,
    width = .9\columnwidth,
    xmajorgrids=true,
    ymajorgrids=true,
    grid style=dashed,
    major tick length = 2,
    minor tick length = 1,
    xtick pos = bottom,
    xticklabel style={
            /pgf/number format/fixed,
            /pgf/number format/precision=0,
            /pgf/number format/fixed zerofill
        },
  },
  % marker
  IronsideM/.style = {
    red!70!black,
    fill=red!50,
  },
  IronsideStaticSafeM/.style = {
    green!70!black,
    postaction={
        pattern={mylines[size=4pt,line width=2pt,angle=45]},
        pattern color=green!70!black,
    },
  },
  IronsideStaticUnsafeM/.style = {
    green!70!black,
    postaction={
        pattern={mylines[size=4pt,line width=2pt,angle=45]},
        pattern color=green!70!black,
    },
  },
  % bar plot
  Ironside/.style = {
    red!80!black,
    fill=red!50,
  },
  IronsideStaticSafe/.style = {
    orange!90!black,
    postaction={
        pattern={mylines[size=4pt,line width=2pt,angle=45]},
        pattern color=orange!50,
    },
  },
  IronsideStaticUnsafe/.style = {
    magenta!80!black,
    postaction={
        pattern={mylines[size=4pt,line width=2pt,angle=135]},
        pattern color=magenta!50,
    },
  },
  IronsideNoCoreMapper/.style = {
    white!00!violet,
    postaction={
        pattern={mylines[size=8pt,line width=4pt,angle=135]},
        pattern color=violet!60,
    },
  },
}

\begin{tikzpicture}[scale=1]
  \begin{groupplot}[CPUUtilGroupPlot]
  % traffic 1 chain 1: p50
  \nextgroupplot[
    LatencyPlot,
    ymode=log,
    ybar,
    ylabel = {p99 Latency [us]},
    ytick = {1, 10, 100, 1000, 10000, 100000},
    ymax = 10000,
    xtick = {100, 200, ..., 500},
    xlabel = {Latency SLO [us]},
    xmax = 550,
  ]
  \addplot [Ironside] table [x expr=\thisrowno{0}, y expr=\thisrowno{4}] {\IronsideTrafficOneChainOne};
  \addplot [IronsideStaticSafe] table [x expr=\thisrowno{0}, y expr=\thisrowno{4}] {\IronsideStaticSafeCoreTrafficOneChainOne};
  \addplot [IronsideStaticUnsafe] table [x expr=\thisrowno{0}, y expr=\thisrowno{4}] {\IronsideStaticUnsafeCoreTrafficOneChainOne};
  \addplot [IronsideNoCoreMapper] table [x expr=\thisrowno{0}, y expr=\thisrowno{4}] {\IronsideNoCoreTrafficOneChainOne};

  % traffic 1 chain 2: p99
  \nextgroupplot[
    LatencyPlot,
    ybar,
    ymode=log,
    ytick = {1, 10, 100, 1000, 10000, 100000},
    xtick = {100, 200, ..., 500},
    xlabel = {Latency SLO [us]},
    xmax = 550,
    ymax = 10000,
  ]
  \addplot [Ironside] table [x expr=\thisrowno{0}, y expr=\thisrowno{4}] {\IronsideTrafficOneChainTwo};
  \addplot [IronsideStaticSafe] table [x expr=\thisrowno{0}, y expr=\thisrowno{4}] {\IronsideStaticSafeCoreTrafficOneChainTwo};
  \addplot [IronsideStaticUnsafe] table [x expr=\thisrowno{0}, y expr=\thisrowno{4}] {\IronsideStaticUnsafeCoreTrafficOneChainTwo};
  \addplot [IronsideNoCoreMapper] table [x expr=\thisrowno{0}, y expr=\thisrowno{4}] {\IronsideNoCoreTrafficOneChainTwo};

  % legend image + text
  \addlegendimage{Ironside};
  \addlegendentry{\Large \sysname};
  \addlegendimage{IronsideStaticSafe};
  \addlegendentry{\Large \sysname static-safe};
  \addlegendimage{IronsideStaticUnsafe};
  \addlegendentry{\Large \sysname static-unsafe};
  \addlegendimage{IronsideNoCoreMapper};
  \addlegendentry{\Large \sysname w/o core-mapper};

  \nextgroupplot[
    LatencyPlot,
    xtick = {100, 200, ..., 500},
    xlabel = {Latency SLO [us]},
    xmax = 550,
    ybar,
    ytick = {10, 20, ..., 50},
    ylabel = {\# of CPU Cores},
    ymax = 40,
    ymin = 0,
  ]
  \addplot [Ironside] table [x expr=\thisrowno{0}, y expr=\thisrowno{2}] {\IronsideTrafficOneChainOne};
  \addplot [IronsideStaticSafe] table [x expr=\thisrowno{0}, y expr=\thisrowno{2}] {\IronsideStaticSafeCoreTrafficOneChainOne};
  \addplot [IronsideStaticUnsafe] table [x expr=\thisrowno{0}, y expr=\thisrowno{2}] {\IronsideStaticUnsafeCoreTrafficOneChainOne};

  \nextgroupplot[
    LatencyPlot,
    xtick = {100, 200, ..., 500},
    xlabel = {Latency SLO [us]},
    xmax = 550,
    ybar,
    ytick = {10, 20, ..., 60},
    ymax = 60,
    ymin = 0,
  ]
  \addplot [Ironside] table [x expr=\thisrowno{0}, y expr=\thisrowno{2}] {\IronsideTrafficOneChainTwo};
  \addplot [IronsideStaticSafe] table [x expr=\thisrowno{0}, y expr=\thisrowno{2}] {\IronsideStaticSafeCoreTrafficOneChainTwo};
  \addplot [IronsideStaticUnsafe] table [x expr=\thisrowno{0}, y expr=\thisrowno{2}] {\IronsideStaticUnsafeCoreTrafficOneChainTwo};

  % \nextgroupplot[
  %   LatencyPlot,
  %   % xmode=log,
  %   xlabel = {Tail Latency [$\mu$s]},
  %   xtick = {1, 10, 100, 1000, 10000, 100000},
  %   xmax = 10000,
  %   xmin = 30,
  %   ylabel = {\# of Cores},
  %   ytick = {5, 10, ..., 50},
  %   ymax = 40,
  %   ymin = 5,
  % ]
  % \addplot [IronsideM] table [x expr=\thisrowno{0}, y expr=\thisrowno{2}] {\IronsideTrafficOneChainOne};
  % \addplot [IronsideStaticSafeM] table [x expr=\thisrowno{0}, y expr=\thisrowno{2}] {\IronsideStaticSafeCoreTrafficOneChainOne};
  % \addplot [IronsideStaticUnsafeM] table [x expr=\thisrowno{0}, y expr=\thisrowno{2}] {\IronsideStaticUnsafeCoreTrafficOneChainOne};

  % \nextgroupplot[
  %   LatencyPlot,
  %   % xmode=log,
  %   xlabel = {Tail Latency [$\mu$s]},
  %   xtick = {1, 10, 100, 1000, 10000, 100000},
  %   xmax = 100000,
  %   xmin = 30,
  %   % ylabel = {\# of Cores},
  %   ytick = {5, 10, ..., 50},
  %   ymax = 40,
  %   ymin = 5,
  % ]
  % \addplot [IronsideM] table [x expr=\thisrowno{0}, y expr=\thisrowno{2}] {\IronsideTrafficOneChainTwo};
  % \addplot [IronsideStaticSafeM] table [x expr=\thisrowno{0}, y expr=\thisrowno{2}] {\IronsideStaticSafeCoreTrafficOneChainTwo};
  % \addplot [IronsideStaticUnsafeM] table [x expr=\thisrowno{0}, y expr=\thisrowno{2}] {\IronsideStaticUnsafeCoreTrafficOneChainTwo};

  \end{groupplot}

  % Chain N texts on right
  \node[] at (3.2cm, -6.9cm) {{{\large CHAIN 1}}};
  \node[] at (10.1cm, -6.9cm) {{{\large CHAIN 2}}};

\end{tikzpicture}

%%% Local Variables:
%%% mode: latex
%%% TeX-master: "../main"
%%% End:
}
    \caption{Comparisons of time-averaged CPU cores used by \sysname and other NFV systems, as a function of achieved p99 latency.}
\label{fig:ablation-core-mapper}
\end{figure}

\begin{table}[!t]
  \small
\centering
\resizebox{0.98\columnwidth}{!}{
\begin{tabular}{|m{1.4cm}|m{1.7cm}|*{3}{>{\centering\arraybackslash}m{1.2cm}}|}
  \hline
  & \textbf{Platform} & \textbf{Avg. Cores} & \textbf{p50} & \textbf{p99} \\
  &  &  & [$\mu$s] & [$\mu$s] \\
  \hline
  \hline
  \multirow{2}{*}{\textbf{Chain 1}} & \textbf{Metron} & 21.8 & 47.7 & 2540  \\
  \cline{2-5} & \textbf{\sysname} & 23.9 & 51.3 & 2126 \\
  \hline
  \multirow{2}{*}{\textbf{Chain 2}} & \textbf{Metron} & 32.1 & 254.5 & 5580  \\
  \cline{2-5} & \textbf{\sysname} & 30.4 & 374.2 & 4567 \\
  \hline
\end{tabular}
}
\caption{Comparisons of end-to-end p50 and p99 latency and the time-averaged CPU core usage by \sysname and Metron. In these experiments, we set \sysname's latency target to 2500 $\mu$s and 5000 $\mu$s respectively. \sysname still produces lower p99 latency results, and has similar CPU core usage.}
\label{t:ironside-high-latency-cost}
\end{table}

% \begin{figure*}[t!]
%     \centering
%     \subfloat[\centering Under backbone traffic]{{
%         \resizebox{0.975\columnwidth}{!}{\input{figs/latency_traffic1.tex}}
%     }}
%     \qquad
%     \subfloat[\centering Under AS traffic]{{
%         \resizebox{0.975\columnwidth}{!}{\input{figs/latency_traffic2.tex}}
%     }}
%     \caption{Comparisons of end-to-end latency achieved by \sysname and other NFV platforms, as a function of latency SLO.}
%     \label{fig:example}
% \end{figure*}

% \begin{figure}[!t]
%     \centering
%     \resizebox{\columnwidth}{!}{\input{figs/core_traffic1.tex}}
%     \vspace{-6mm}
%     \caption{Comparisons of time-averaged CPU cores used by \sysname and other NFV systems, as a function of achieved p99 latency.}
% \label{fig:core_traffic1}
% \end{figure}

\subsection{Ablation Study: The Core Mapper}
\label{subsec:eval-core-mapper}

\sysname's core mapper reacts to bursts at each dedicated core on a worker and uses flow and packet counts to make flow-to-core mapping decisions. To understand the importance of these design choices, we compare \sysname to 3 different variants: 1) \textit{static-safe}: always uses the smallest packet count regardless of the number of flow arrivals; 2) \textit{static-unsafe}: always uses the largest packet count regardless of the number of flow arrivals; 3) \textit{w/o core mapper}: a \sysname variant that disables core mapper.

\parae{Results.} \figref{fig:ablation-core-mapper} demonstrates that only \sysname is able to provide sub-ms latency SLOs when using its core mapper or uses \textit{static-safe} core mapper. Without core mapper, \sysname is unable to achieve sub-ms latency. When using \textit{static-unsafe} core mapper, \sysname achieves sub-ms p99 latency, but violates SLOs in some cases. The \textit{static-safe} variant can result in a smaller p99 latency in most cases, but may use up to 13\% more CPU cores, compared to \sysname's design so we use the latter (\secref{s:absorb-bursts:-core}).

\subsection{Ablation Study: The Server Mapper}
\label{subsec:eval-server-mapper}

\sysname's server mapper finds the fewest dedicated cores on a \sysname worker in response to persistent load shifts. As with the core mapper, we explore two variants: \textit{static-safe} that allocates cores based on the lowest packet count, and \textit{static-unsafe} allocates cores based on the highest packet count.

\parae{Results.} \figref{fig:ablation-server-mapper} shows that all three designs are able to meet sub-ms p99 latency SLOs with comparable CPU core usage for chain 1. For chain 2, \sysname's design dominates both in terms of latency and CPU core usage, indicating that, in some settings, the server mapper must consider flow and packet rates for efficient SLO-compliance.

\parab{The necessity of boost-mode.} Instead of reconfiguring RSS tables under persistent overload, \sysname's boost-mode uses the dedicated core to drain packet queues, and migrates NF processing to other cores (\secref{s:core-effic-serv}). To show the importance of having boost-mode, we compare \sysname against two different variants: 1) \textit{w/o boost-mode}: \sysname variant that disables the boost-mode; 2)  \textit{w/ on-demand invocations}: instead of enabling boost-mode, this variant invokes the server mapper's RSS-based flow-to-core remapping immediately when a persistent overload presents.

\parae{Results.} \figref{fig:ablation-boost-mode} shows that enabling boost-mode is crucial for achieving sub-ms tail latency SLOs. The two other \sysname variants can only result in ms-scale tail latency; the first because it cannot handle load changes at second timescales and the second because of the overhead of re-programming RSS tables.

\begin{figure}[!t]
    \centering
    \resizebox{\columnwidth}{!}{% traffic 1 / chain 1
\pgfplotstableread[row sep=\\,col sep=&]{
SLO & Loss  & Core & p50  & p99    \\
100 & 0.006 & 33.6 & 28.8 & 89.4   \\
200 & 0.006 & 30.7 & 32.1 & 174.4  \\
300 & 0.005 & 30.1 & 30.7 & 238.8  \\
400 & 0.010 & 30.5 & 34.5 & 254.4  \\
500 & 0.005 & 30.7 & 31.1 & 298.2  \\
}\IronsideTrafficOneChainOne

\pgfplotstableread[row sep=\\,col sep=&]{
SLO & Loss  & Core & p50  & p99      \\
100 & 0.038 & 33.51 & 31.4 & 111.2   \\
200 & 0.038 & 29.46 & 30.5 & 147.2   \\
300 & 0.059 & 30.35 & 30.9 & 137.1   \\
400 & 0.055 & 29.94 & 32.2 & 155.5   \\
500 & 0.053 & 30.07 & 36.5 & 332.4   \\
}\IronsideStaticSafeServerTrafficOneChainOne

\pgfplotstableread[row sep=\\,col sep=&]{
SLO & Loss  & Core & p50  & p99      \\
100 & 0.044 & 32.73 & 33.6 & 169.9   \\
200 & 0.047 & 30.97 & 36.7 & 230.9   \\
300 & 0.076 & 29.79 & 38.5 & 256.4   \\
400 & 0.097 & 29.19 & 37.0 & 373.3   \\
500 & 0.085 & 29.48 & 41.2 & 384.9   \\
}\IronsideStaticUnsafeServerTrafficOneChainOne

% traffic 1 / chain 2
\pgfplotstableread[row sep=\\,col sep=&]{
SLO & Loss  & Core & p50  & p99     \\
100 & 0.004 & 56.0 & 18.0 & 102.4   \\
200 & 0.015 & 41.9 & 22.8 & 180.3   \\
300 & 0.006 & 42.8 & 24.8 & 197.2   \\
400 & 0.017 & 40.3 & 24.9 & 210.3   \\
500 & 0.007 & 40.6 & 26.8 & 270.5   \\
}\IronsideTrafficOneChainTwo

\pgfplotstableread[row sep=\\,col sep=&]{
SLO & Loss  & Core   & p50  & p99      \\
100 & 0.013 & 59.6   & 0    & 102.3   \\
200 & 0.006 & 45.7   & 0    & 173.5   \\
300 & 0.015 & 44.97  & 0    & 244.3   \\
400 & 0.003 & 45.11  & 0    & 235.8   \\
500 & 0.010 & 43.11  & 0    & 305.4   \\
}\IronsideStaticSafeServerTrafficOneChainTwo

\pgfplotstableread[row sep=\\,col sep=&]{
SLO & Loss  & Core   & p50  & p99      \\
100 & 0.013 & 56.25  & 0    & 107.4   \\
200 & 0.006 & 43.99  & 0    & 181.6   \\
300 & 0.015 & 44.77  & 0    & 258.1   \\
400 & 0.003 & 45.46  & 0    & 256.2   \\
500 & 0.010 & 45.52  & 0    & 336.7   \\
}\IronsideStaticUnsafeServerTrafficOneChainTwo

\pgfplotsset{
  % single bar in legend
  /pgfplots/ybar legend/.style={
    /pgfplots/legend image code/.code={
      \draw[##1,/tikz/.cd, yshift=-0.3em]
      (7 mm, 0 mm) rectangle (11pt, 0.75em);},
  },
  CPUUtilGroupPlot/.style = {
    group/group name = cpuutils,
    group/group size = 2 by 2,
    group/xlabels at = edge bottom,
    group/ylabels at = edge left,
    group/vertical sep = 35pt,
  },
  LatencyPlot/.style = {
    font = \large,
    height = .7\columnwidth,
    width = .9\columnwidth,
    bar width=.23cm,
    tick pos = left,
    % xmajorgrids=true,
    ymajorgrids=true,
    major tick length = 2,
    xlabel near ticks,
    ylabel near ticks,
    xtick align = outside,
    ytick align = outside,
    xlabel shift = -3 pt,
    legend columns = 3,
    legend cell align = left,
    legend style = {
       font = \normalsize,
       fill=none,
       at = {(0.98, 1.24), anchor = south},
      /tikz/every even column/.append style={column sep=0.15cm},
    }
  },
  CPUUtilPlot/.style = {
    font = \large,
    legend columns = 4,
    legend cell align = left,
    legend style = {
       font = \footnotesize,
       at = {(0.98, 1.2), anchor = south},
      /tikz/every even column/.append style={column sep=0.10cm},
    },
    height = .5\columnwidth,
    width = .9\columnwidth,
    xmajorgrids=true,
    ymajorgrids=true,
    grid style=dashed,
    major tick length = 2,
    minor tick length = 1,
    xtick pos = bottom,
    xticklabel style={
            /pgf/number format/fixed,
            /pgf/number format/precision=0,
            /pgf/number format/fixed zerofill
        },
  },
  % marker
  IronsideM/.style = {
    red!70!black,
    fill=red!50,
  },
  IronsideStaticSafeM/.style = {
    green!70!black,
    postaction={
        pattern={mylines[size=4pt,line width=2pt,angle=45]},
        pattern color=green!70!black,
    },
  },
  IronsideStaticUnsafeM/.style = {
    green!70!black,
    postaction={
        pattern={mylines[size=4pt,line width=2pt,angle=45]},
        pattern color=green!70!black,
    },
  },
  % bar plot
  Ironside/.style = {
    red!80!black,
    fill=red!50,
  },
  IronsideStaticSafe/.style = {
    orange!90!black,
    postaction={
        pattern={mylines[size=4pt,line width=2pt,angle=45]},
        pattern color=orange!50,
    },
  },
  IronsideStaticUnsafe/.style = {
    magenta!80!black,
    postaction={
        pattern={mylines[size=4pt,line width=2pt,angle=135]},
        pattern color=magenta!50,
    },
  },
}

\begin{tikzpicture}[scale=1]
  \begin{groupplot}[CPUUtilGroupPlot]
  % traffic 1 chain 1: p99
  \nextgroupplot[
    LatencyPlot,
    xtick = {100, 200, ..., 500},
    xlabel = {Latency SLO [us]},
    xmax = 550,
    ybar,
    ytick = {0, 50, ..., 500},
    ylabel = {p99 Latency [us]},
    ymax = 400,
    ymin = 0,
  ]
  \addplot [Ironside] table [x expr=\thisrowno{0}, y expr=\thisrowno{4}] {\IronsideTrafficOneChainOne};
  \addplot [IronsideStaticSafe] table [x expr=\thisrowno{0}, y expr=\thisrowno{4}] {\IronsideStaticSafeServerTrafficOneChainOne};
  \addplot [IronsideStaticUnsafe] table [x expr=\thisrowno{0}, y expr=\thisrowno{4}] {\IronsideStaticUnsafeServerTrafficOneChainOne};

  % traffic 1 chain 2: p99
  \nextgroupplot[
    LatencyPlot,
    xtick = {100, 200, ..., 500},
    xlabel = {Latency SLO [us]},
    xmax = 550,
    ybar,
    ytick = {0, 50, ..., 500},
    ymax = 400,
    ymin = 0,
  ]
  \addplot [Ironside] table [x expr=\thisrowno{0}, y expr=\thisrowno{4}] {\IronsideTrafficOneChainTwo};
  \addplot [IronsideStaticSafe] table [x expr=\thisrowno{0}, y expr=\thisrowno{4}] {\IronsideStaticSafeServerTrafficOneChainTwo};
  \addplot [IronsideStaticUnsafe] table [x expr=\thisrowno{0}, y expr=\thisrowno{4}] {\IronsideStaticUnsafeServerTrafficOneChainTwo};

  % legend image + text
  \addlegendimage{Ironside};
  \addlegendentry{\Large \sysname};
  \addlegendimage{IronsideStaticSafe};
  \addlegendentry{\Large \sysname static-safe};
  \addlegendimage{IronsideStaticUnsafe};
  \addlegendentry{\Large \sysname static-unsafe};

  \nextgroupplot[
    LatencyPlot,
    xtick = {100, 200, ..., 500},
    xlabel = {Latency SLO [us]},
    xmax = 550,
    ybar,
    ytick = {10, 20, ..., 50},
    ylabel = {\# of CPU Cores},
    ymax = 40,
    ymin = 0,
  ]
  \addplot [Ironside] table [x expr=\thisrowno{0}, y expr=\thisrowno{2}] {\IronsideTrafficOneChainOne};
  \addplot [IronsideStaticSafe] table [x expr=\thisrowno{0}, y expr=\thisrowno{2}] {\IronsideStaticSafeServerTrafficOneChainOne};
  \addplot [IronsideStaticUnsafe] table [x expr=\thisrowno{0}, y expr=\thisrowno{2}] {\IronsideStaticUnsafeServerTrafficOneChainOne};

  \nextgroupplot[
    LatencyPlot,
    xtick = {100, 200, ..., 500},
    xlabel = {Latency SLO [us]},
    xmax = 550,
    ybar,
    ytick = {10, 20, ..., 60},
    ymax = 60,
    ymin = 0,
  ]
  \addplot [Ironside] table [x expr=\thisrowno{0}, y expr=\thisrowno{2}] {\IronsideTrafficOneChainTwo};
  \addplot [IronsideStaticSafe] table [x expr=\thisrowno{0}, y expr=\thisrowno{2}] {\IronsideStaticSafeServerTrafficOneChainTwo};
  \addplot [IronsideStaticUnsafe] table [x expr=\thisrowno{0}, y expr=\thisrowno{2}] {\IronsideStaticUnsafeServerTrafficOneChainTwo};

  \end{groupplot}

  % Chain N texts on right
  \node[] at (3.2cm, -6.95cm) {{{\large CHAIN 1}}};
  \node[] at (10.1cm, -6.95cm) {{{\large CHAIN 2}}};

\end{tikzpicture}

%%% Local Variables:
%%% mode: latex
%%% TeX-master: "../main"
%%% End:
}
    \caption{Comparisons of time-averaged CPU cores used by \sysname and other NFV systems, as a function of achieved p99 latency.}
\label{fig:ablation-server-mapper}
\end{figure}

\begin{figure}[!t]
    \centering
    \resizebox{\columnwidth}{!}{% traffic 1 / chain 1
\pgfplotstableread[row sep=\\,col sep=&]{
SLO & Loss  & Core & p50  & p99    \\
100 & 0.006 & 33.6 & 28.8 & 79.4   \\
200 & 0.006 & 30.7 & 32.1 & 174.4  \\
300 & 0.005 & 30.1 & 30.7 & 238.8  \\
400 & 0.010 & 30.5 & 34.5 & 254.4  \\
500 & 0.005 & 30.7 & 31.1 & 298.2  \\
}\IronsideTrafficOneChainOne

\pgfplotstableread[row sep=\\,col sep=&]{
SLO & Loss  & Core & p50  & p99      \\
100 & 0.024 & 21.8 & 47.7 & 2540.6   \\
200 & 0.024 & 21.8 & 47.7 & 2540.6   \\
300 & 0.024 & 21.8 & 47.7 & 2540.6   \\
400 & 0.024 & 21.8 & 47.7 & 2540.6   \\
500 & 0.024 & 21.8 & 47.7 & 2540.6   \\
}\IronsideNoBoostTrafficOneChainOne

\pgfplotstableread[row sep=\\,col sep=&]{
SLO & Loss  & Core & p50  & p99      \\
100 & 0.062 & 43.3 & 30.8 & 3050.8   \\
200 & 0.057 & 43.3 & 39.1 & 3020.7   \\
300 & 0.056 & 43.3 & 38.8 & 3014.7   \\
400 & 0.052 & 43.2 & 39.9 & 2988.4   \\
500 & 0.044 & 41.7 & 53.7 & 2994.2   \\
}\IronsideOnDemandTrafficOneChainOne

% traffic 1 / chain 2
\pgfplotstableread[row sep=\\,col sep=&]{
SLO & Loss  & Core & p50  & p99     \\
100 & 0.004 & 56.0 & 18.0 & 102.4   \\
200 & 0.015 & 41.9 & 22.8 & 180.3   \\
300 & 0.006 & 42.8 & 24.8 & 197.2   \\
400 & 0.007 & 40.3 & 24.9 & 210.3   \\
500 & 0.007 & 40.6 & 26.8 & 270.5   \\
}\IronsideTrafficOneChainTwo

\pgfplotstableread[row sep=\\,col sep=&]{
SLO & Loss  & Core & p50  & p99        \\
100 & 0.044 & 32.04 & 254.5 & 5580.3   \\
200 & 0.044 & 32.04 & 254.5 & 5580.3   \\
300 & 0.044 & 32.04 & 254.5 & 5580.3   \\
400 & 0.044 & 32.04 & 254.5 & 5580.3   \\
500 & 0.044 & 32.04 & 254.5 & 5580.3   \\
}\IronsideNoBoostTrafficOneChainTwo

\pgfplotstableread[row sep=\\,col sep=&]{
SLO & Loss  & Core & p50  & p99      \\
100 & 0.126 & 76.7 & 34.0 & 5458.9   \\
200 & 0.133 & 78.3 & 34.7 & 5465.4   \\
300 & 0.132 & 75.5 & 34.0 & 5451.6   \\
400 & 0.137 & 77.1 & 34.5 & 5472.7   \\
500 & 0.135 & 77.2 & 34.2 & 5454.8   \\
}\IronsideOnDemandTrafficOneChainTwo

\pgfplotsset{
  % single bar in legend
  /pgfplots/ybar legend/.style={
    /pgfplots/legend image code/.code={
      \draw[##1,/tikz/.cd, yshift=-0.3em]
      (7 mm, 0 mm) rectangle (11pt, 0.75em);},
  },
  CPUUtilGroupPlot/.style = {
    group/group name = cpuutils,
    group/group size = 2 by 1,
    group/xlabels at = edge bottom,
    group/ylabels at = edge left,
    group/vertical sep = 25pt,
  },
  CPUUtilPlot/.style = {
    font = \large,
    height = .7\columnwidth,
    width = .9\columnwidth,
    bar width=.23cm,
    tick pos = left,
    % xmajorgrids=true,
    ymajorgrids=true,
    major tick length = 2,
    xlabel near ticks,
    ylabel near ticks,
    xtick align = outside,
    ytick align = outside,
    legend columns = 2,
    legend cell align = left,
    legend style = {
       font = \normalsize,
       fill=none,
       at = {(0.98, 1.31), anchor = south},
      /tikz/every even column/.append style={column sep=0.15cm},
    }
  },
  Ironside/.style = {
    red!80!black,
    fill=red!50,
  },
  IronsideNoBoost/.style = {
    orange!90!black,
    postaction={
        pattern={mylines[size=4pt,line width=2pt,angle=45]},
        pattern color=orange!50,
    },
  },
  IronsideOnDemand/.style = {
    white!00!violet,
    postaction={
        pattern={mylines[size=8pt,line width=4pt,angle=135]},
        pattern color=violet!60,
    },
  },
}

\begin{tikzpicture}[scale=1]
  \begin{groupplot}[CPUUtilGroupPlot]
  % traffic 1 chain 1: p50
  \nextgroupplot[
    CPUUtilPlot,
    ymode=log,
    ylabel = {p99 Latency [us]},
    xtick = {100, 200, ..., 500},
    ytick = {1, 10, 100, 1000, 10000, 100000},
    xmax = 550,
    ymax = 10000,
    ybar,
  ]
  \addplot [Ironside] table [x expr=\thisrowno{0}, y expr=\thisrowno{4}] {\IronsideTrafficOneChainOne};
  \addplot [IronsideNoBoost] table [x expr=\thisrowno{0}, y expr=\thisrowno{4}] {\IronsideNoBoostTrafficOneChainOne};
  \addplot [IronsideOnDemand] table [x expr=\thisrowno{0}, y expr=\thisrowno{4}] {\IronsideOnDemandTrafficOneChainOne};

  % traffic 1 chain 2: p99
  \nextgroupplot[
    CPUUtilPlot,
    ymode=log,
    xtick = {100, 200, ..., 500},
    ytick = {1, 10, 100, 1000, 10000, 100000},
    xmax = 550,
    ymax = 10000,
    ybar,
  ]
  \addplot [Ironside] table [x expr=\thisrowno{0}, y expr=\thisrowno{4}] {\IronsideTrafficOneChainOne};
  \addplot [IronsideNoBoost] table [x expr=\thisrowno{0}, y expr=\thisrowno{4}] {\IronsideNoBoostTrafficOneChainTwo};
  \addplot [IronsideOnDemand] table [x expr=\thisrowno{0}, y expr=\thisrowno{4}] {\IronsideOnDemandTrafficOneChainTwo};

  % legend image + text
  \addlegendimage{Ironside};
  \addlegendentry{\Large \sysname};
  \addlegendimage{IronsideNoBoost};
  \addlegendentry{\Large \sysname w/o boost-mode};
  \addlegendimage{Quadrant};
  \addlegendentry{\Large \sysname w/ on-demand invoc.};

  \end{groupplot}

  % Chain N texts on right
  % \node[] at (3.4cm, -7.5cm) {{{\large CHAIN 1}}};
  % \node[] at (11.4cm, -7.5cm) {{{\large CHAIN 2}}};
  % \node[] at (19.4cm, -7.5cm) {{{\large CHAIN 1}}};
  % \node[] at (27.4cm, -7.5cm) {{{\large CHAIN 2}}};

  % \node[] at (7.2cm, -8.1cm) {{{\Large (a) Backbone Traffic}}};
  % \node[] at (23.2cm, -8.1cm) {{{\Large (b) AS Traffic}}};

\end{tikzpicture}

%%% Local Variables:
%%% mode: latex
%%% TeX-master: "../main"
%%% End:
}
    \caption{Comparisons of time-averaged CPU cores used by \sysname and other NFV systems, as a function of achieved p99 latency.}
% \vspace{1mm}
\label{fig:ablation-boost-mode}
\end{figure}

\subsection{Analyzing Scheduling Overheads}
\label{subsec:eval-overhead}

\parab{Enqueue and dequeue overhead.} In \sysname, a dedicated core pulls the packet from a NIC queue and enqueues the packet to a software queue (either owned by the core or an auxiliary core) before it is processed. The packet will be dequeued and being processed later. We now quantify this enqueue/dequeue overhead (\sysname uses a lock-less queue).

\parae{Results.} By computing the avg. cycle cost of 10k enqueue and dequeue operations, we find: \textbf{enqueue} adds 40 cycles / batch; \textbf{dequeue} adds 32 cycles / batch. The overhead is small compared to the packet processing time of an NF chain. 

\parab{Core-remapping overhead.} At the end of each short epoch, \sysname's core mapper invokes a core re-mapping to balance workloads on software queues and recruit auxiliary cores when necessary. This core re-mapping includes decision and enforcement processes that scan all packets in a software queue (owned by a dedicated core).

\parae{Results.} \tabref{t:core-remapping-cost} shows benchmark results on the execution time of the core re-mapping processing under different packet queue lengths. As shown, even for more than 1k packets, this core re-mapping process can be finished in less than 5 $\mu$s. In reality, \sysname can control software queue length with its server mapper design, and expect much fewer packets in a software queue. This overhead is relatively small for 100s-$\mu$s latency SLOs. To extend \sysname for tighter SLOs, re-design might be required and we have left this to future work.

\begin{table}[t]
\centering
\resizebox{0.95\columnwidth}{!}{
  \begin{tabular}{|m{2.5cm}|*{2}{>{\centering\arraybackslash}m{2.4cm}}|}
  \hline
  \textbf{{Pkt queue length}} & \textbf{\small{Core re-mapping process}} [cycles] & \textbf{\small{Core re-mapping process}} [$\mu$s] \\
  % & [cycles] & [$\mu$s] \\
  \hline
  \hline
  \textbf{128} & 4.2k & 1.5 \\
  \hline
  \textbf{256} & 7.1k & 2.6 \\
  \hline
  \textbf{512} & 7.7k & 2.8 \\
  \hline
  \textbf{1024} & 12.2k & 4.4 \\
  % \hline
  % 2048 & 22.4k & 8.0 \\
  \hline
\end{tabular}
}
\caption{The average execution time of \sysname's core mapper's core re-mapping process, as a function of packet queue length. This process runs at the end of each short epoch, and is lightweight (i.e. $<5 \mu$s to re-map more than 1k packets).}
\label{t:core-remapping-cost}
\end{table}

\section{Related Work}
\label{sec:related}

\sysname is unique in enabling microsecond SLOs for NF chains with high CPU core efficiency in a cluster setting, but draws inspiration from several areas in the literature.

% CPU scheduling: tail latency, cpu core efficiency
% ZygOS, Shenego, Caladan, Breakwater, RackSched, 
\parab{CPU scheduling.} A line of recent work has explored microsecond scale RPCs for datacenter applications. ZygOS and RackSched~\cite{sosp17-zygos, osdi20-racksched} study scheduling policies for serving RPC requests in a server/rack setting,  Arachne~\cite{osdi18-arachne} allows applications to manage cores based on loads,  Shenango~\cite{nsdi19-shenango} co-locates latency-sensitive and batch processing applications for higher CPU efficiency, and Caladan~\cite{osdi20-caladan} mitigates interference for co-located tasks. A recent follow-up~\cite{nsdi22-scheduling_policy} explores optimal combinations of load balancing and scaling for microsecond-scale RPCs. \sysname is inspired by this body of work on the trade-off between latency and efficiency but for a different setting: NFV, whose flow-affinity requirement for stateful NFs results in qualitatively different solutions. 

% Fastpass: making packet scheduling decisions for low queueing delay
% Queueing study: packet scheduling in the network
% (please add more reference if you know some
\parab{Packet scheduling.} Fastpass~\cite{sigcomm14-fastpass} uses the centralized packet-level scheduling to achieve a `zero-queue' network. \sysname optimizes for per-worker NIC queues rather than switch queues, and is subject to traffic with a higher packet and flow rate, which requires a hierarchical design rather than a centralized design.

% NFV
% e2, opennetvm, netbricks, safebricks, galleon, edgeos
% opennf, statelessnf, s6
% execution: resq, nfvnice
% scaling: metron, galleon, snf
% Hardware offloading: why they are orthogonal to us 

\parab{NFV.} Prior work falls into four groups. NFV frameworks~\cite{sigcomm14-opennf,nsdi14-netvm,sosp15-e2,osdi16-netbricks,nsdi18-safebricks,socc22-quadrant,atc20-edgeos} solve problems in developing, chaining and isolating NFs. While \sysname's focus is different, it can use these mechanisms to chain and isolate NFs. For performance optimization, a series of papers propose mechanisms on efficiently managing NF states for stateful NFs~\cite{sigcomm16-openbox,nsdi17-stateless,nsdi-18-s6}, on optimizing resource allocations when NFs run on top of a shared resource pool~\cite{nsdi18-resq}, and on minimizing losses when chaining NFs across cores~\cite{sigcomm17-nfvnice}. Researchers have been exploring approaches for offloading NF processing to hardware, including SmartNICs, OpenFlow and PISA switches, GPU, FPGA etc.~\cite{eurosys15-nba,socc17-uno,nsdi18-metron,conext20-lemur,osdi20-pigasus}. Finally, towards a practical NFV deployment, the community has been looking for deploying NFV in a cluster to serve dynamic traffic, which involves load balancing and scaling. These proposals~\cite{nsdi18-metron,conext19-rsspp,infocom22-dyssect,socc20-snf,socc22-quadrant} are relevant, but they are insufficient for achieving microsecond SLOs for stateful NF chains (as shown in \secref{sec:evaluation}).

\section{Conclusion}
\label{sec:conclusion}

In this paper, we identify a key problem in NFV: achieving microsecond-scale tail latency SLOs with high CPU core efficiency. Our trace analysis reveals two types of traffic bursts in realistic traffic, and shows that NFV systems must handle bursts properly to achieve the objective.
We posit that NFV systems need to adapt quickly to transient traffic bursts. Then, we propose a suite of scaling mechanisms: the core mapper that is able to absorb transient bursts at each CPU core, and the server mapper that assigns and migrates flows to as few CPU cores as possible to ensure high CPU efficiency.
Our NFV system, \sysname, is able to meet sub-ms tail latency SLOs while other today's NFV systems cannot.

\clearpage
\bibliography{references.bib}

\newpage

\appendix

\renewcommand\thefigure{A.\arabic{figure}}
\renewcommand\thesection{\Alph{section}}
\renewcommand\thetable{A.\arabic{table}}
\setcounter{section}{0}
\setcounter{figure}{0}
\setcounter{table}{0}

\section*{Appendix}
\label{sec:appendix}

% \lipsum[1-4]
\section{The Server Mapper's Heuristic}
\label{sec:server-mapper-heuristic}

\sysname's server mapper is designed to ensure high core efficiency by continuously re-allocating flows to the minimum number of dedicated cores to process the current traffic demand.

\begin{figure}[h!]
    \includegraphics[width=0.95\columnwidth]{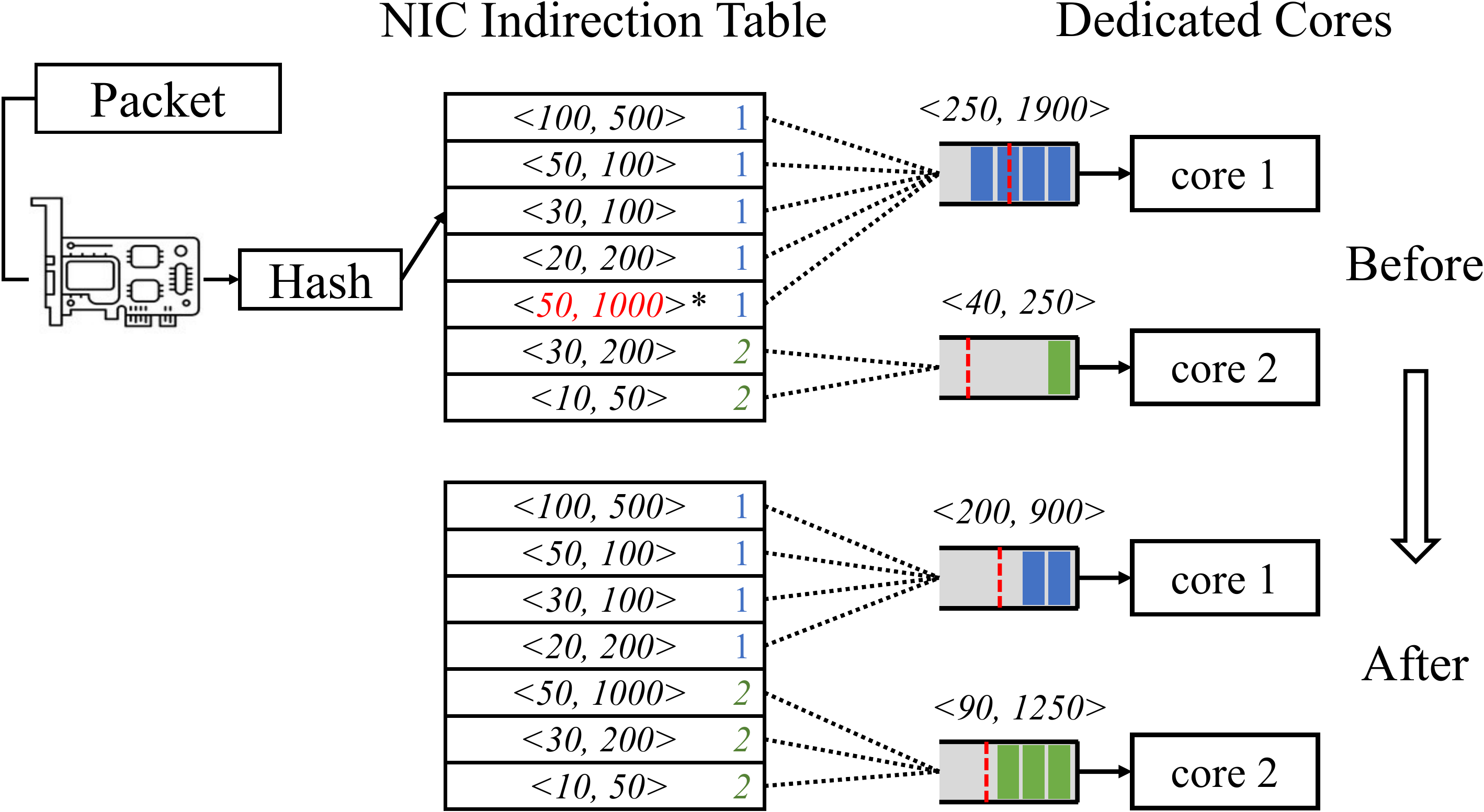}
\caption{The server mapper leverages RSS, and updates the NIC's indirection table to apply bucket-to-core mappings. For each decision interval, it tracks the flow count and packet rate $<f,r>$ for each RSS bucket, corresponding to one entry in the NIC's table. At the end of an interval, the server mapper decides the number of dedicated cores at a server: it finds overloaded cores, whose loads exceed the core's capacity, moves the minimum set of buckets from them to other cores, and then tries to reclaim cores. In the above case, one bucket is migrated from core 1 to core 2 to avoid overloading.}
\label{fig:server-mapper}
\end{figure}

\sysname uses the NIC RSS technique to map flow aggregates to cores quickly. Rather than modeling it as a constrained mixed-integer linear problem (MILP), we propose a fast greedy heuristics for traffic re-distribution. \figref{fig:server-mapper} provides an example of re-mapping RSS buckets to cores to demonstrate this algorithm.

%%% Local Variables:
%%% mode: latex
%%% TeX-master: "main"
%%% End:

\end{document}